\documentclass[reprint,amsmath,amsfonts,amsgen,amssymb,aps, pra, twocolumn,nofootinbib]{revtex4-1}

\usepackage{dcolumn}
\usepackage[normalem]{ulem}
\usepackage{mathtools}
\usepackage{braket}
\usepackage{footmisc}
\usepackage{hyperref}
\usepackage{siunitx}
\usepackage{cases}
\usepackage{svg}

\usepackage{bm}     
\usepackage{float}

\usepackage{xcolor}
\usepackage{booktabs}   

\begin{document}

\title{Light-harvesting enhanced by quantum ratchet states}

\author{Nicholas Werren}
\thanks{These authors contributed equally.}
\email{n.werren@hw.ac.uk}
\address{SUPA, Institute of Photonics and Quantum Sciences, Heriot-Watt University, Edinburgh, EH14 4AS, United Kingdom}

\author{William M. Brown \textsuperscript{*}}
\address{SUPA, Institute of Photonics and Quantum Sciences, Heriot-Watt University, Edinburgh, EH14 4AS, United Kingdom}

\author{Erik M. Gauger}
\email{e.gauger@hw.ac.uk}
\address{SUPA, Institute of Photonics and Quantum Sciences, Heriot-Watt University, Edinburgh, EH14 4AS, United Kingdom}

\date{\today}

\begin{abstract}
We consider bio-inspired ring systems as photovoltaic circuits to explore the advantage of `optical ratcheting', a process whereby the arrangement of coupled optical dipoles enables delocalised excitonic states that are protected against radiative decay whilst permitting the absorption of further photons.
We explore how the performance of a ratcheting antenna scales with system size when excitons are incoherently or coherently extracted from the antenna to an associated trap site.
In both instances we also move to the polaron frame in order to more closely model realistic systems where the coupling to vibrational modes can generally be assumed to be strong.
We find a multifaceted and nuanced dependence of the predicted performance on an interplay between geometrical arrangement, extraction mechanism, and vibrational coupling strength. Certain regimes support substantial performance improvements in the power extracted per site. 

\end{abstract}


\maketitle

\section{Introduction}
Understanding the mechanics of light-harvesting in nature is a subject of intense interest owing to the powerful implications a complete understanding of photosynthesis would have for improving technologies of solar energy conversion
\cite{Scholes11, Alharbi15, Hu18, Tomasi19,  Blazquez19}.
A significant obstacle facing the development of such technology is substantial energy loss through inefficiency, e.g.~through conversion of light energy to heat and loss via re-radiation.
In the context of conventional solid-state light-harvesting, this difficulty is captured by the Shockley–Queisser limit which places a theoretical maximum efficiency for single-band gap solar cells \cite{Shockley61}.
A key contribution to this limit comes from the condition of detailed balance, which states that the absorption of light by an optical system must be accompanied by an equivalent emission (and vice versa). 
It has been demonstrated theoretically \cite{Scully10, Scully11}, and recently also experimentally \cite{Bittner14, Trebbia22}, that detailed balance can be broken. 
Developing alternative or complementary ways in which to achieve this under conditions more closely resembling natural solar irradiation is a profoundly relevant subject of contemporary research.

Conventional silicon $p$-$n$ junctions have achieved efficiencies exceeding 26\% \cite{Andreani19}, approaching the Shockley–Queisser limit which restricts this technology to 29\% \cite{Tiedje84, Green84}. Molecular (organic) platforms offer an attractive alternative to $p$-$n$ junctions and opportunities for boosting light-harvesting efficiency which are not available to conventional $p$-$n$ junctions, however, it remains an open challenge to fulfil their potential in practical designs.
Owing to their promise of relatively cheap cost and simple fabrication organic photovoltaics (OPVs) \cite{Xue10, Zheng10} present themselves as a viable means of harvesting solar power on a large scale.
Recent advances have been demonstrated to provide a power conversion efficiency of up to 17\% \cite{Cui19, Rui20}.
New developments in the fabrication of OPVs further extends the possibilities for utilising new polymer structures for efficient power conversion \cite{Cindy14, Seok15}.
As such, a key obstacle yet to be overcome by OPVs is achieving power conversion efficiencies resembling, or even exceeding, those of inorganic solar cells.

Utilising more complex molecular structures could provide novel mechanisms through which the obstacles facing conventional single-band gap solar cells can be overcome.
For instance, the interference between multiple dipoles in a shared environment can prevent radiative transitions between certain energy levels; 
such optically inaccessible states are known as dark states.
By introducing a vibrational environment these optically dark states can become preferentially populated, resulting in so-called dark state protection \cite{Creatore13, Zhang15, Zhang16, Fruchtman16, Rouse19}. 
In this way, dark states provide a mechanism through which radiative loss is mitigated by suppressing exciton recombination and thereby protecting excited population from decay.
Such mitigation relies on a careful architecture in the energetic structure of a system \cite{Brown19, Davidson20}.
Therefore, dark state protection offers a means for longer range transport in artificial and natural light-harvesting systems \cite{Mattioni21, Davidson22}.

Generally, the geometric arrangement of constituent optical dipoles has a profound effect upon exhibited collective characteristics of quantum systems, such as sub- \cite{MorenoCardoner19, Turschmann19} and super-radiance \cite{Holzinger21, Masson2022}, both of which have been verified through experimental observation \cite{Guerin16, Solano2017}.
Of particular interest are ring systems, which exhibit a number of collective phenomena.
Two examples are the enhanced emission \cite{Holzinger20, Holzinger22} and absorption of these ring systems \cite{MorenoCardoner22} through dark state protection.

As we show here, ring configurations that are (loosely) inspired by biological photosynthetic antennae can also give rise to the conditions appropriate for a recurring optical ratcheting effect beyond simple dark state protection. Importantly, the requisite ring structures (of different sizes) can artificially engineered and exhibit certain coherent features \cite{OSullivan11}. 
Notably, examples of nanoscale rings that can be fabricated using porphyrin polymers \cite{Raymond08, Zheng16, Cai17, Richert17, Judd20} include diporphyrins (i.e.~dimer structures) \cite{Chang83, Ouyang09}, as well as cyclic trimer \cite{Metselaar08, GilRamirez10}, quadmer (tetramer) \cite{Maeda20} and pentamer \cite{Liu16, Yu18} ring configurations.
Further instances of ring structures could emerge from the organization of chromophores through so-called DNA origami \cite{Hemmig16, Ketterer18, Dey21}.

\textit{Optical ratcheting} is a mechanism where the coupling to optical and vibrational baths is structured such that detailed balance is broken through the organisation of bright and dark states \cite{Higgins17}.
For ratcheting to occur, the energy separation between dark and bright states must fall into the vibrational spectrum of the phonon bath.
Dark states are thereby populated through thermal vibrational relaxation. 
The dark states can then absorb a second photon and move into a bright state in the next excitation manifold.

\begin{figure*}[htbp]
    \includegraphics[height=10.0 cm, angle=0]{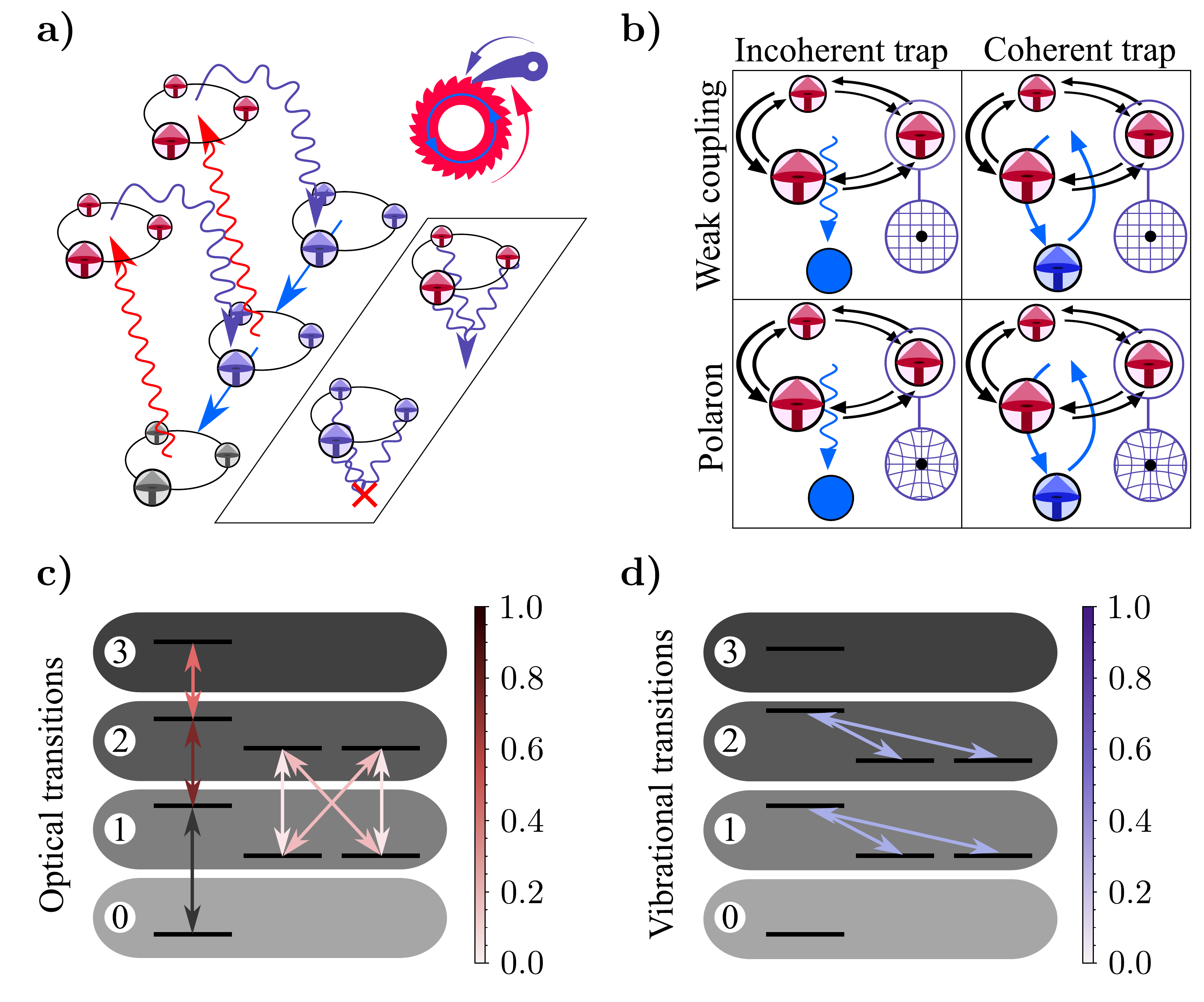}
    \abovecaptionskip=0pt
    \caption{Optical ratcheting with ring antennae. \textbf{a)} The basic mechanism of ``ratcheting" - the transitioning of a system from bright (red) to dark (purple) state where optical emission is prevented but absorption is still possible. The slanted boxed inset indicates how optical decay from bright states is enhanced by constructive interference, whereas dark states are protected by destructive interference between the decay paths of the individual emitters. \textbf{b)} A diagrammatic depiction of the different models, serving as a legend for our results figures. Panels \textbf{c)} and \textbf{d)} capture the optical and vibrational transitions, respectively, for a trimer antenna. The arrows show the allowed transitions between ring eigenstates whilst their colour denotes relative (normalised) strengths. The (white) encircled black numbers indicate the relevant excitation manifold of the antenna.} 
    \label{fig:RatchetDiag1}
\end{figure*}

In this Article, we study the phenomenon of such optical ratcheting for the first time using physically realistic and appropriate models for molecular systems. Specifically, we incorporate strong vibrational couplings and extraction with a coherent trap into our description of bio-inspired ring systems.
Only the simplified example of an incoherent trap extracting from an antenna weakly interacting with its vibrational environment -- see the top left of panel b) in Fig.~\ref{fig:RatchetDiag1} -- has been previously investigated \cite{Higgins17, Zhang15}.
Starting in this configuration, we here address the question of both the requisite, and the optimal ring size for ratcheting. We find trimer antennae are the simplest and closest to ideal ratcheting antenna unit across a broad range of parameter space. We proceed to explore the power harvested in more rigorous models, as presented in the remaining sections of Fig.~\ref{fig:RatchetDiag1}b, where vibrational interactions are treated in the polaron frame, and energy extraction is supported via Hamiltonian dipolar coupling terms.
Our study highlights the practicality,  flexibility, and robustness of optical ratcheting, demonstrating light-harvesting performance improvements that could be translated into enhancing the efficiency of future photocell technologies.

\section{Model} \label{sec:mod}

We consider an antenna system composed of a ring of $N$ optical dipoles which each interact with one another and are coupled to their surrounding environments, namely optical and vibrational baths.
The state of the total system is described by the system-bath density matrix, $\rho(t)=\rho_{S}(t)\otimes\rho_{B}(t)$, where the density matrix of the bath, $\rho_{B}(t)=\rho_{vib}(t)\otimes\rho_{opt}(t)$, captures the optical and vibrational baths.
We now outline the microscopic Hamiltonian model of the ring antenna, which acts as the foundation of all of our photocell models.
Firstly, we make the dipole approximation and assume our molecular chromophores can be approximated as pointlike optical dipoles.
Interactions between closely spaced optical dipoles manifest as (resonant F\"orster) dipolar coupling terms in a (antenna) system Hamiltonian \cite{Mukai99, Scholes00, Warshel86}, and can, for example, be derived in the framework of the many-body quantum optical master equation \cite{Varada92, Curutchet17}.
Therefore, for our antenna system, we first consider a ring of two-level optical dipoles described by the following Hamiltonian:
\begin{equation} \label{eq:Hs}
    \hat{H_{S}} = \omega_{s} \sum^{N}_{j=1}  \hat{\sigma}_{j}^{z} + \sum^{N}_{j,k=1} J_{j,k}(\textbf{r}_{j,k}) \left(\hat{\sigma}_{j}^{+} \hat{\sigma}_{k}^{-} + \hat{\sigma}_{j}^{-} \hat{\sigma}_{k}^{+} \right),
\end{equation}
where the first term is the bare Hamiltonian of the ring sites, all with equal transition frequency, $\omega_{s}$, and where $\hat{\sigma}_{j}^{z}$ is the usual Pauli $z$-operator acting on site $j$. 
The second term captures the resonant F\"orster interactions, where relative dipole orientations and separations are accounted for by the prefactor $J_{j,k}(\textbf{r}_{j,k})$:
\begin{multline}
\label{eq:fcoup}
    J_{j,k}(\textbf{r}_{j,k}) = \frac{1}{4\pi \epsilon_{0}} \bigg( \frac{\textbf{d}_{j} \cdot  \textbf{d}_{k}}{|\textbf{r}_{j,k}|^{3}} \\
    - \frac{3 (\textbf{r}_{j,k} \cdot \textbf{d}_{j})(\textbf{r}_{j,k} \cdot \textbf{d}_{k})}{|\textbf{r}_{j,k}|^{5}} \bigg)~,
\end{multline}
which is a geometrical factor between the $j$th and $k$th dipoles of oscillator strength $|\textbf{d}_{j}|$ and $|\textbf{d}_{k}|$ separated by distance $\textbf{r}_{j,k}$.
The paired raising and lowering operators, $\hat{\sigma}_{j}^{+}$ and $\hat{\sigma}_{k}^{-}$, in Eqn.~\ref{eq:Hs} capture the movement of an exciton between the $j$th and $k$th sites.
For this study, we shall assume all optical dipoles are aligned perpendicular to the plane of the ring. Note that deviations from this arrangement can lead to interesting effects other than ratcheting, as, e.g., explored for tilted dipoles in Ref.~\cite{Brown19}.
Our choice here also contrasts with the arrangement found in light-harvesting complexes such as LHI and LHII, where dipoles tend to align tangentially around the ring \cite{Savage96, Hu01}.

The dipoles composing the ring interact collectively with a shared optical bath, and as the size of the ring system is much smaller\footnote{Having the dipoles closely spaced is required in order to have sizeable F\"orster interaction terms.} than the wavelength of relevant photons, $\lambda \approx  2 \pi c/\omega_s$, we can assume all dipoles couple with the same position-independent phase to the electric field.
The couplings of these dipoles to an optical field are then captured by the optical interaction Hamiltonian \cite{Breuer02Book},
\begin{equation} \label{eq:HIopt}
    \hat{H}_{I,opt} = \sum^{N}_{j=1}\textbf{d}_{j} \hat{\sigma}_{j}^{x} \otimes \sum_{p} f_{p} (\hat{a}_{p} + \hat{a}_{p}^{\dag})~,
\end{equation}
where $\hat{a}_{p}^{(\dag)}$ are the annihilation (creation) operators for the $p$th optical mode, and $f_{p}$ is the coupling strength of the $p$th mode.
We further assume the dipoles are each coupled to identical local phonon modes.
In the context of molecules such phonons capture intramolecular vibrations, as well as interactions with the surrounding local environment.
Thus, we consider a vibrational interaction Hamiltonian of the form:
\begin{equation} \label{eq:HIvib}
    \hat{H}_{I,vib} = \sum^{N}_{j=1}\hat{\sigma}_{j}^{z} \otimes \sum_{q} g_{q} \left(\hat{b}_{j, q} + \hat{b}_{j, q}^{\dag} \right)~,
\end{equation}
where $g_{q}$ and $\hat{b}_{j, q}^{(\dag)}$ are the coupling strength and annihilation (creation) operators for the $q$th phonon mode with the $j$th dipole, respectively.
Lastly, the optical and phonon environmental modes are governed by the following Hamiltonian:
\begin{equation}
    \hat{H}_{B} = \sum_{p} \omega_{p} \hat{a}^{\dag}_{p} \hat{a}_{p} + \sum_{j, q} \tilde{\omega}_{j, q} \hat{b}^{\dag}_{j, q} \hat{b}_{j, q}~,
\end{equation}
where $\omega_{p}$ are the frequencies of the $p$th photon modes, and $\tilde{\omega}_{j, q}$ are the frequencies of the $q$th phonon modes corresponding to the $j$th dipole.
The total Hamiltonian is then given by,
\begin{equation} \label{eq:Htot1}
     \hat{H} = \hat{H}_{S} + H_{I} + \hat{H}_{B}~,
\end{equation}
where $H_{I}=\hat{H}_{I,vib} +  \hat{H}_{I,opt} $.
To summarise, this Hamiltonian governs the dynamics of a ring of parallel optical dipoles in interaction with each other, a shared optical bath, and individual local phonon baths as shown in Fig.~\ref{fig:RatchetDiag1}.
By making the Born approximation, the evolution of the system is then captured by performing a perturbative expansion to second order in the environmental couplings by the following generic form of quantum master equation \cite{Breuer02Book}:
\begin{multline} \label{eq:BRME}
    \frac{d }{dt}\rho_{S}(t) = -\frac{i}{\hbar} \left[\hat{H}_{S}, \rho_{S}(t)\right] \\
    - \frac{1}{\hbar^{2}} \int^{\infty}_{0} \text{Tr}_{B}\left[\hat{H}_{I}(t),\left[\hat{H}_{I}(t-s), \rho_{S}(t) \otimes \rho_{B} \right]\right]ds~.
\end{multline}
We now proceed to describe the different specific models considered in this work.

\subsection{Antenna model}
\textit{Weak Vibrational Coupling --- }
As the interaction terms Eqns.~\eqref{eq:HIopt} and \eqref{eq:HIvib} take the form $\hat{H}=\hat{A}\otimes\hat{B}$, and by assuming that the interaction with the environment is weak, we derive the non-secular Redfield dissipators \cite{Breuer02Book}, $\tilde{\mathcal{D}}_{opt}$ and $ \tilde{\mathcal{D}}_{vib}$ for the optical and vibrational environments, respectively, describing the dissipative processes associated with each.
The second term in Eqn.~\eqref{eq:BRME} is reduced to pairwise combinations of eigenoperators $\hat{A}_{n}$ in the eigenbasis:
\begin{multline}
    \tilde{\mathcal{D}} = \sum_{n,m} \Gamma_{nm}\left(\omega\right)  \left(A_{m}\left( \omega_{m}\right) \rho(t) A^{\dag}_{n}\left( \omega_{n}\right) \right. \\
    \left. -  A^{\dag}_{n}\left( \omega_{n}\right) A_{m}\left( \omega_{m}\right) \rho(t) + \mathrm{H.c.} \right)~,
\end{multline}
where $\mathrm{H.c.}$ is the Hermitian conjugate of the preceding terms, and the environmental contribution to the interaction is captured by the rate $\Gamma_{nm}(\omega)$, describing a transition between the $n$th and $m$th eigenstates,
\begin{equation}
    \Gamma_{nm}(\omega) = \int^{\infty}_{0} e^{i\omega s} \braket{\hat{B}^{\dag}_{n}(t) \hat{B}_{m}(t-s)} ds~.
\end{equation}
We assume that each environment is in a thermal state, thereby finding the following expression for the rates:
\begin{equation}
    \Gamma_{nm}(\omega) = \frac{1}{2}\gamma_{nm}(\omega) + i S_{nm}(\omega)~,
\end{equation}
where $S_{nm}$ is a typically small energy shift which we can neglect.
The remaining contribution to the rate,
\begin{equation} \label{eq:ispecdens}
    \gamma_{nm}(\omega) = J(\omega)N(\omega)~,
\end{equation}
includes the spectral density, $J(\omega)$, and
\begin{numcases}
{N(\omega)=}
  (1 + n(\omega)), & \( \omega \geq 0\) \nonumber \\
   n(\omega), & \( \omega < 0 \)   \nonumber
\end{numcases}
where the Bose-Einstein distribution 
\begin{equation}
    n(\omega)=\frac{1}{e^{\hbar \omega \beta}-1}~
\end{equation}
with $\beta=1/k_{B}T$, is calculated using the temperature of the corresponding bath.
Therefore, for the optical dissipator, these terms take the form
\begin{equation}
     \gamma^{opt}_{nm}(\omega) = \frac{\omega_{nm}^{3}}{\omega_{s}^{3}}  \kappa_{opt}N(\omega)~,
\end{equation}
where $ \kappa_{opt}$ is the spontaneous decay rate of a single, isolated dipole.
We have assumed that the temperature of the optical bath is $T_{opt}=5800$~K \cite{WurfelBook, Zhang16, Fruchtman16}.

For the vibrational bath we assume a structureless Ohmic spectral density \cite{Knox02, Toutounji02}, which behaves similar to Drude-Lorentz at high temperature, and which has been used to model excitonic transfer in light-harvesting complexes \cite{Fassioli12, Kreisbeck14}.
We thereby construct the following rate:
\begin{equation}
     \gamma^{vib}_{nm}(\omega) = \kappa_{vib }\frac{\omega}{\bar{\omega}_{vib}} \left( n(\omega) + 1 \right)~,
\end{equation}
where $\kappa_{vib}$ represents a characteristic phonon timescale, here chosen to be on the order of picoseconds, and $\bar{\omega}_{vib}$ is the average phonon transition frequency\footnote{This assumption ensures the necessary hierarchy of optical and vibrational baths for ratcheting.}.
Note that in order for ratcheting to function in the outlined system, the timescale of phonon relaxation must be the fastest in the hierarchy of relevant rates, but given such a hierarchy its specific value does not significantly affect the results. 
Therefore, the full quantum master equation for the density matrix, $\rho$, of the antenna system described, takes the form:
\begin{equation}
\label{eq:QMEincohweak}
    \frac{d}{d t} \rho_{S} = -i \left [\hat{H}_{S}, \rho \right] + \tilde{\mathcal{D}}_{opt} +  \tilde{\mathcal{D}}_{vib}~.
\end{equation}

\textit{Polaron Frame --- }
In the case of a stronger interaction between antenna and vibrational environment, as can be expected in many typical molecular systems, the model explored thus far is not sufficient.
Such a situation is then better described following a polaron transformation, which affects the phonon bath in such a way as to completely remove the phonon interaction terms \cite{Gross82, Nazir16}.
The transformation is captured by the following unitary operator:
\begin{equation}
    \hat{U}_{B} = \exp\left[\sum^{N}_{j=1}\hat{\sigma}_{j}^{z} \otimes \sum_{q} g_{q} \left(\hat{b}_{j, q} + \hat{b}_{j, q}^{\dag} \right) \right]~.
\end{equation}
The transformed antenna system Hamiltonian includes an effective rescaling of the transition frequencies and hopping terms:
\begin{multline} \label{eq:HsPol2}
    \hat{H}^{\prime}_{S}  = \hat{U}_{B} \hat{H}_{S} \hat{U}^{\dag}_{B}= \sum^{N}_{j=1} \omega^{s \prime}_{j} \hat{\sigma}_{j}^{z} \\
    + \sum^{N}_{j\neq k} J_{j,k}(\textbf{r}_{j,k}) \braket{\hat{B}^{\pm}}^{2}\left(\hat{\sigma}_{j}^{+} \hat{\sigma}_{k}^{-} + \hat{\sigma}_{j}^{-} \hat{\sigma}_{k}^{+}\right)~,
\end{multline} 
where the frequency $\omega^{s \prime}_{j} = \omega^{s}_{j} - \Lambda$ is shifted by a renormalisation energy,
\begin{equation}
    \Lambda=\int^\infty_{0} \frac{J(\omega)}{\omega} d\omega~,
\end{equation}
and we have introduced the displacement terms,
\begin{equation}
    \hat{B}^{\pm}_{j} = \prod_{q} \exp\left[\pm\left(\frac{g_{j,q}}{\omega_{q}} \hat{b}^{\dag}_{q, j} - \frac{g^{*}_{j,q}}{\omega_{q}} \hat{b}_{q, j} \right)\right]~.
\end{equation}
In the continuum limit the expectation value for the displacement terms becomes \cite{Nazir16}:
\begin{multline}
    \braket{\hat{B}^{\pm}_{j}} \equiv \braket{\hat{B}} \\
    = \exp\left[-\frac{1}{2} \int^{\infty}_{0}d\omega\frac{ J(\omega)}{\omega} \coth\left( \frac{\beta \omega}{2}\right)\right]~.
\end{multline}
Importantly, as a consequence of the off-diagonal elements of $\hat{H}_{S}$, the transformation introduces a further contribution from the F\"orster coupling which acts as an additional interaction term:
\begin{multline} \label{eq:HsPol2b}
    \hat{H}_{I, coup} = \sum^{N}_{j\neq k} J_{j,k}(\textbf{r}_{j,k}) \left(\hat{B}^{+}_{j}\hat{B}^{-}_{k} - \braket{\hat{B}}^{2}\right)\hat{\sigma}^{+}_{j}\hat{\sigma}^{-}_{k} \\
    + \hat{\sigma}^{-}_{j}\hat{\sigma}^{+}_{k}\left(\hat{B}^{-}_{j}\hat{B}^{+}_{k} - \braket{\hat{B}}^{2}\right)~,
\end{multline} 
leading to a further dissipative process described by
\begin{multline}
    \tilde{\mathcal{D}}^{\prime}_{coup} = - \frac{1}{\hbar^{2}} \int^{\infty}_{0} \text{Tr}_{B}\left[\hat{H}_{I, coup}(t),\right. \\
    \left. \left[\hat{H}_{I, coup}(t-s), \rho_{S}(t) \otimes \rho_{B} \right]\right]ds \\
    = \sum_{n,m} \Gamma^{vib}_{nm}\left(\omega\right)  \left(A_{m}\left( \omega_{m}\right) \rho(t) A^{\dag}_{n}\left( \omega_{n}\right) \right. \\
    \left. -  A^{\dag}_{n}\left( \omega_{n}\right) A_{m}\left( \omega_{m}\right) \rho(t) + \mathrm{H.c.} \right)~.    
\end{multline}
As in the weak coupling case, the dissipative phonon rates $\Gamma^{vib}_{nm}$ in the polaron frame are captured by bath correlations, now appropriately displaced. 
For an ensemble of two-level systems these amount to the following pairwise combinations of `raising' and `lowering' operators:
\begin{equation}
    \braket{\hat{B}^{\pm}_{ j}(s)\hat{B}^{\pm}_{ k}(0)} = \braket{\hat{B}}^{2}e^{-\phi(s)}~,
\end{equation}
\begin{equation}
    \braket{\hat{B}^{\pm}_{j}(s)\hat{B}^{\mp}_{k}(0)} = \braket{\hat{B}}^{2}e^{\phi(s)}~,
\end{equation}
and
\begin{equation}
    \braket{\hat{B}^{\pm} _{j}(s)}\braket{\hat{B}^{\pm}_{k }(0)} = \braket{\hat{B}^{\pm}_{j }(s)}\braket{\hat{B}^{\mp}_{k}(0)} = \braket{\hat{B}}^{2}~,
\end{equation}
where
\begin{multline}
    \phi(s)=\int^{\infty}_{0}d\omega\frac{J(\omega)}{\omega^{2}}\bigg[\cos(\omega s) \coth\left(\frac{\beta \omega}{2}\right) \\
    - i\sin(\omega s)\bigg]~.
\end{multline}
The polaron transformation requires a superohmic spectral density, $J(\omega)$.
For simplicity we therefore introduce
\begin{equation}
\label{eq:specdens2}
    J(\omega)=\frac{\lambda \omega^{3}}{2 \omega^{3}_{c}} e^{-\frac{\omega}{\omega_{c}}}~,
\end{equation}
with reorganisation energy, $\lambda$, and a cut-off frequency, $\omega_{c}$ \cite{Sowa17, Brown19} \footnote{The primary influence of the spectral density at high temperatures is the system renormalisation, and therefore the reorganisation energy is more relevant than the specific shape of density \cite{Knox02, Toutounji02}, hence our choice of rates in this limit.}.

We can now consider the dissipative terms, which will now include the interaction captured by Eqn.~\eqref{eq:HsPol2b}, in the polaron transformed Redfield master equation. 
We also introduce the transformed optical interaction Hamiltonian,
\begin{multline} \label{eq:HIopt2}
    \hat{H}^{\prime}_{I,opt} = \sum^{N}_{j=1}\textbf{d}_{j}\left( \hat{B}^{+}_{j}\hat{\sigma}_{j}^{+} + \hat{B}^{-}_{j}\hat{\sigma}_{j}^{-} \right) \\
    \otimes \sum_{k} f_{k} (\hat{a}_{k} + \hat{a}_{k}^{\dag})~,
\end{multline}
which contains both optical and vibrational operators.
The resulting dissipator is described by,
\begin{multline}
    \tilde{\mathcal{D}}^{\prime}_{opt} = - \frac{1}{\hbar^{2}} \int^{\infty}_{0} \text{Tr}_{B}\left[\hat{H}^{\prime}_{I,opt}(t),\right. \\
    \left. \left[\hat{H}^{\prime}_{I,opt}(t-s), \rho_{S}(t) \otimes \rho_{B} \right]\right]ds \\
    = P_{vib} \sum_{n,m} \Gamma^{opt}_{nm}\left(\omega\right)  \left(A_{m}\left( \omega_{m}\right) \rho(t) A^{\dag}_{n}\left( \omega_{n}\right) \right. \\
    \left. -  A^{\dag}_{n}\left( \omega_{n}\right) A_{m}\left( \omega_{m}\right) \rho(t) + \mathrm{H.c.} \right)~.
\end{multline}
Here, $P_{vib}$ are weightings contributed by the vibrational environment to the optical term. The difference of timescales between photon and phonon processes allows us to separate phonon correlations from the optical rates, evaluating the former at zero time \cite{Nazir16} (although this approximation can be relaxed if necessary \cite{Rouse22}).
As such, the phonon contribution $P_{vib}$ reduces to the following prefactors:
\begin{equation}
    \braket{\hat{B}^{\pm }_{j}(0)\hat{B}^{\pm }_{k}(0)} = \braket{\hat{B}}^{4}~,
\end{equation}
and
\begin{equation}
    \braket{\hat{B}^{\mp }_{j}(0)}\braket{\hat{B}^{\pm }_{k}(0)} = 1~.
\end{equation}
We therefore find the following quantum master equation for the antenna system:
\begin{equation} \label{eq:QMEincohstrong}
    \frac{d}{d t} \rho_{S} = -i \left [\hat{H}^{\prime}_{S}, \rho \right] + \tilde{\mathcal{D}}_{opt}^{\prime} + \tilde{\mathcal{D}}^{\prime}_{coup}~,
\end{equation}
where the Hamiltonian and optical dissipator are now appropriately transformed.
With the antenna dynamics fully outlined it is possible to introduce a trap, and thereby describe the extraction processes with which we can investigate the advantages of optical ratcheting.

\subsection{Trap (extraction) model}
We shall employ two different models for extracting excitons from the ring antenna for subsequent conversion to useful energy.
The first is the conventional \textit{incoherent extraction} model, wherein a trap is introduced through phenomenological extraction terms \cite{Higgins17}, whereas the second is a \textit{coherent extraction} model, in which the trap is introduced physically as an additional dipole that is coherently coupled to the antenna ring system via a dipole-dipole Hamiltonian term.

In either case, the trap is treated as a two-level system, with excited state, $\ket{e}$, and ground state, $\ket{g}$.
As a result, in both models we consider a joint density matrix of antenna system and trap, $\rho=\rho_{s}\otimes\rho_{t}$.
The framework of the quantum heat engine (QHE) \cite{Scully11, Dorfman13} allows for the calculation of current and voltage.
Here, the introduction of the photo-excited ring system is regarded as a photovoltaic circuit into which the trap is integrated.
An important characteristic of these cycles is that the rate of extraction can produce a bottleneck \cite{Caruso12, Blankenship14Book}.
As we will show, in this limit, and when designing optimal architectures regarding the ratio of antennae vs `reaction centres' \cite{Higgins17}, ratcheting offers significant efficiency gains. 

\textit{Incoherent Extraction --- }
For incoherent extraction, the energy splitting of the trap, $\omega_{t}$, is chosen such that the trap is degenerate with the lowest energy eigenstate of the first excited manifold of the ring, where population will gather under the ratcheting process.
The decay rate of the trap $\Gamma^{\downarrow}_{t}$ represents a hypothetical (electric) load, whereby the exciton is converted to useful work from the system.
We construct the appropriate quantum master equation by including $\bar{\mathcal{D}}_{X}$ and $\bar{\mathcal{D}}^{\downarrow}_{t}$ in Eqn.~\eqref{eq:QMEincohweak}, where these Lindblad dissipators terms capture incoherent exciton extraction and trap decay, respectively.
The extraction term takes the standard Lindblad form
\begin{equation}
    \bar{\mathcal{D}}_{X}=\Gamma_{X} \left( \hat{X}\rho\hat{X}^{\dag} - \frac{1}{2}\{\hat{X}^{\dag}\hat{X}, \rho\} \right) ~,
\end{equation}
where the rate $\Gamma_{X}$ represents the rate of exciton extraction via incoherent hopping and the operator $\hat{X}$ describes this process,
\begin{equation}
    \hat{X} = \hat{\sigma}^{-}\otimes\hat{\sigma}_{t}^{+}~,
\end{equation} 
in which the trap is extracting from a single antenna dipole in the site basis \cite{Higgins17}.
Alternative extraction configurations are possible e.g., extraction from multiple sites or a single energy eigenstate.

\footnote{Note that our choice of (incoherent) extraction model here will efficiently extract from all ring eigenstates, and we reserve the complications of more realistic energy selectivity to our coherent trap model.}.
Similarly, for the trap decay term,
\begin{equation}
\bar{\mathcal{D}}^{\downarrow}_{t}=\Gamma^{\downarrow}_{t} \left(\hat{\sigma}^{-}_{t}\rho \hat{\sigma}^{+}_{t} - \frac{1}{2}\{\hat{\sigma}^{+}_{t}\hat{\sigma}^{-}_{t}, \rho\} \right) ~,
\end{equation}
where $\sigma^{-}_{t}$ ($\sigma^{+}_{t}$) is the lowering (raising) operator acting on the trap dipole, and $\Gamma^{\downarrow}_{t}$ is the rate of trap decay.
Finally, we reach the complete quantum master equation for incoherent extraction in the polaron frame which is described by
Eqn.~\eqref{eq:QMEincohstrong}, where we introduce additional dissipator terms $\bar{\mathcal{D}}^{\prime}_{X}$ and $\bar{\mathcal{D}}^{\downarrow \prime}_{t}$. Note that populations remain unaltered under the polaron transformation, such that extracting incoherently in the site basis is consistent in both frames considered here.

\textit{Exciton-exciton annihilation --- }
Many (natural) organic systems experience non-radiative de-excitation when two excitons meet and interact, a mechanism known as exciton-exciton annihilation (EEA). Typically, this involves a transition from two excitations on adjacent sites to a single higher excited state on one site that is followed by fast non-radiative relaxation\cite{Gaididei80, May14}. 
Rather than going into microscopic detail, we here introduce EEA through an operationally motivated decay process at rate $\Gamma_{EEA}$ and described by the dissipative term
\begin{multline}
\bar{\mathcal{D}}_{EEA}=\Gamma_{EEA}\sum^{N}_{n=2} \left(\hat{\varsigma}^{-}_{(n)}\rho \hat{\varsigma}^{+}_{(n)} - \frac{1}{2}\{\hat{\varsigma}^{+}_{(n)}\hat{\varsigma}^{-}_{(n)}, \rho\} \right) ~,
\end{multline}
where $\hat{\varsigma}^{-}_{(n)} = \sum_{i,j} \ket{\xi_{n-1}^{j}}\bra{\xi_{n}^{i}}$ describes the decay of population from the $n$-th excitation manifold into the eigenstates of the $(n-1)$-th manifold $\ket{\xi_{n-1}^{j}}$. Here, $\ket{\xi_{n}^{i}}$ labels all the eigenstates in the $n$-th manifold via its index $i$. Further, $\Gamma_{EEA}$ is the rate of the associated EEA mechanism. Using a single `decay' operator between adjacent manifolds (as opposed to a number of independent ones) is unproblematic as our steady states correspond to an incoherent mixture of eigenstates, and we have checked both models deliver identical results. Note that in our results this term is only present where explicitly stated.

\textit{Coherent Extraction --- }
The coherent model of extraction describes a trap dipole coherently extracting from an antenna, see the top right of Panel b) in Fig.~\ref{fig:RatchetDiag1}. 
We model this by including the trap as an additional dipole, such that an additional Hamiltonian term for the coherent trap $\hat{H}_{t}$ is included in the unitary dynamics:
\begin{equation} \label{eq:Ht}
    \hat{H_{t}} = \omega_{t}\hat{\sigma}^{z}_{t} + \sum^{N}_{j=1} J_{j,t}(\textbf{r}_{j,t}) \left(\hat{\sigma}_{j}^{+}\otimes \hat{\sigma}_{t}^{-} + \hat{\sigma}_{j}^{-}\otimes \hat{\sigma}_{t}^{+} \right).
\end{equation}
The first term in Eqn.~\eqref{eq:Ht} captures the energy splitting, $\omega_t$, of the trap.
Importantly, the second term describes an additional dipole-dipole coupling which links the trap with every other site on the ring, in a geometry-dependent way.
The quantum master equation is described by Eqn.~\eqref{eq:QMEincohweak} with the addition of the coherent trap $\hat{H}_{t}$ and the corresponding decay rate $\bar{\mathcal{D}}^{\downarrow}_{t}$.
Notably, under coherent extraction, the energy splitting of the trap, $\omega_{t}$, is treated as a free variable in order to correctly tune the coherent trap with the antenna. 
The resulting polaron master equation is therefore captured by Eqn.~\eqref{eq:QMEincohstrong}, with the inclusion of a coherent trap described in the appropriate frame $\hat{H}^{\prime}_{t}$.

Lastly, in our adopted QHE formalism \cite{Scully10, Dorfman13, Creatore13}, for all models of extraction the current is obtained with the following equation:
\begin{equation} \label{eq:curr}
    I = e \Gamma^{\downarrow}_{t} \braket{\rho_{t,e}}_{SS}~,
\end{equation}
where $\braket{\rho_{t,e}}_{SS}$ is the steady state population of the excited trap state.
Under this approach the potential difference is derived from the population of trap states in relation to the trap's governing thermal distribution,
\begin{equation} \label{eq:QHE}
    eV = \hbar \omega_{t} + k_{B} T_{vib} \ln{\left[\frac{\braket{\rho_{t,e}}_{SS}}{\braket{\rho_{t,g}}_{SS}}\right]}~,
\end{equation}
where $k_{B}$ is Boltzmann’s constant, $T_{vib}$ is the phonon bath temperature, and the expectation value, $\braket{\rho_{t,g}}_{SS}$, is the steady state population of the trap's ground state.
Straightforwardly, the first term in Eqn.~\eqref{eq:QHE} describes the bare transition energy of the trap.
The second term accounts for differences in the trap populations from the thermal distribution, ensuring both thermodynamic consistency \cite{Gelbwaser17}, and providing a mechanism for varying $V$ via its dependence on $\Gamma^{\downarrow}_{t}$.
This will give us access to $IV$ curves and allow us to find the optimal power point, $P_{\rm{max}} = \max{(I V)}$.

\section{Results} \label{sec:res}

\begin{table}[H]
\begin{center}
\begin{tabular}{*6l}  
\toprule
\textbf{Parameter} & \textbf{Symbol} & \textbf{Default value} &&&  \\
\midrule
Dipole ring radius & $R$  & 1.5~\si{nm}  \\ 
Atomic transition frequency  & $\omega^{s}_{j}$  & 2~\si{eV}  \\ 
Spontaneous decay rate & $\kappa_{opt}$ & $2.5$~\si{ns^{-1}}\\ 
Phonon relaxation rate & $\kappa_{vib}$  & $1$~\si{ps^{-1}} \\
Photon bath temperature & $T_{opt}$ & $5800$~\si{K} \\
Phonon bath temperature & $T_{vib}$ & $300$~\si{K} \\
Exciton annihilation rate & $\Gamma_{EEA}$ & $25$~\si{ns^{-1}}\\
\hline
\bottomrule
\end{tabular}
\caption{Default model parameters, similar to those used in \cite{Creatore13, Zhang15, Zhang16, Higgins17}.
The values provided are used for all calculations unless parameters appear on axes of plots, or explicitly stated otherwise.}
\label{tab:params1}
\end{center}
\end{table}

\subsection{Incoherent extraction} \label{sec:resincoh}
 
We proceed by calculating the steady state solutions of the quantum master equations for an incoherent trap coupled to an antenna, described by a ring of $N$ dipoles, in a weak vibrational coupling regime or the polaron frame [Eqns.~\eqref{eq:QMEincohweak} or \eqref{eq:QMEincohstrong}].
We quantify the ratcheting advantage by calculating the difference between the power extracted \textit{per site} from ring systems using the following equations:
\begin{equation} \label{eq:powdiff}
   \Delta  \bar{P}_{NM} = \frac{P_{N}}{N} - \frac{P_{M}}{M}~,
\end{equation}
and
\begin{equation} \label{eq:powdiffp}
   \Delta  \bar{P}^{\%}_{NM} = \frac{1}{N}\frac{P_{N}}{P_{M}} - \frac{1}{M}~,
\end{equation}
where $N$ and $M$ are the number of sites constituting the ring system.
We then use the current and voltage from the quantum heat engine model, Eqns.~\eqref{eq:curr} and \eqref{eq:QHE}, to explore the performance of these ring systems as energy harvesters.
When comparing trimer with dimer and quadmer systems we use the measure defined by Eqns.~\eqref{eq:powdiff} and \eqref{eq:powdiffp}, which provides a means of comparing the power extracted from two different ring systems whilst accounting for the increase in current, which additional absorbing antenna dipoles necessarily provide.
Whilst in the following we focus on the generated power, ratcheting can be understood as primarily boosting electric current. 
We therefore provide the associated current plots in Appendix~\ref{sec:curr}.
Unless explicitly stated otherwise, results are calculated using the parameters from Table \ref{tab:params1}.

\begin{figure*}[htbp]
    \includegraphics[height=12 cm, angle=0]{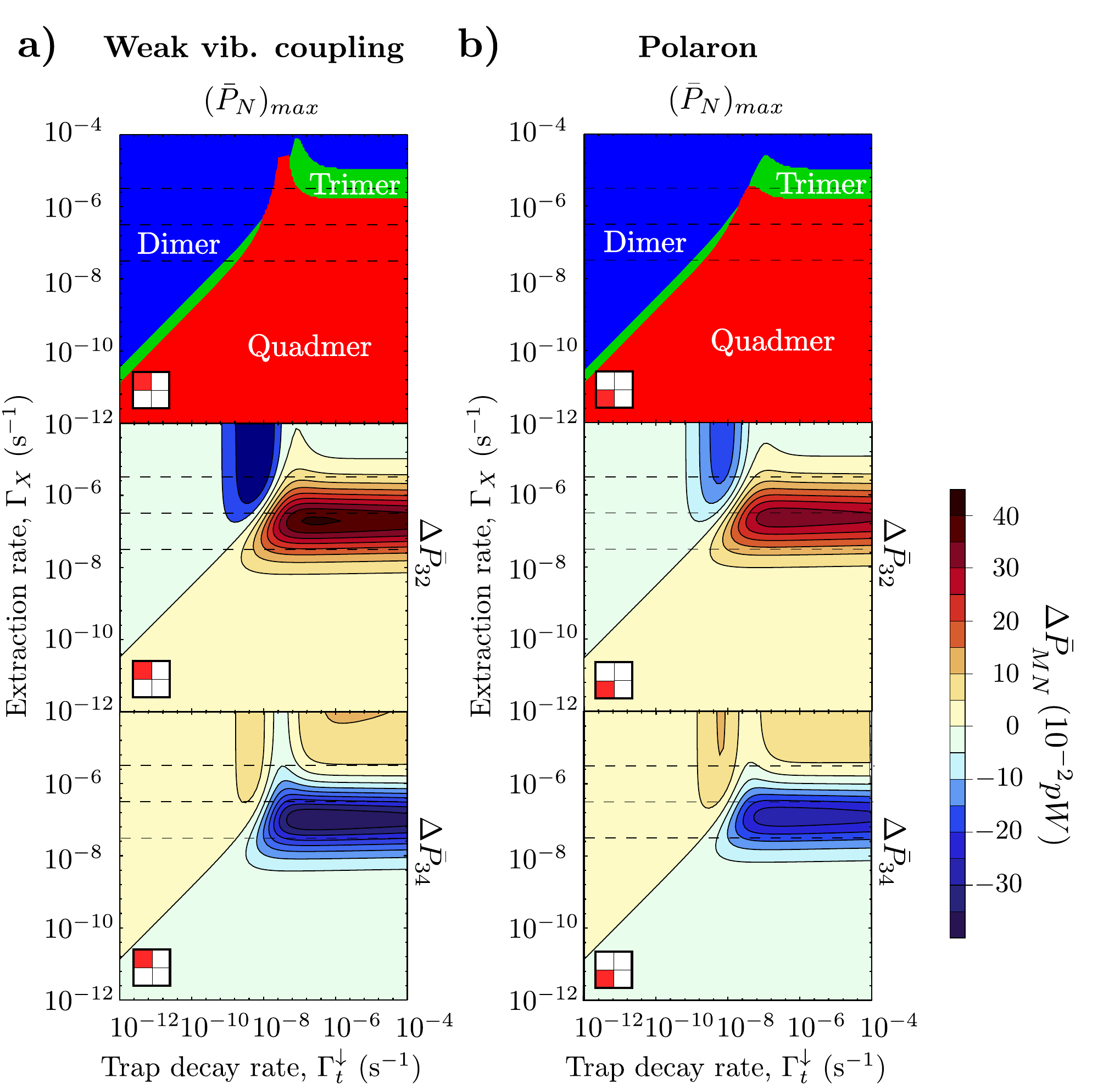}
    \abovecaptionskip=0pt
    \caption{
    Power generated from a ring system by an incoherent trap in the weakly coupled and polaron regimes on the left and right hand side respectively.
    In the top panels we highlight the optimal ring system, these differences are explicitly shown in the middle and bottom panels for the trimer relative to the dimer and quadmer respectively.
    Values of $\Gamma_{X}$ used for later plots are indicated by dashed lines.
    Excluding those labelled, parameters are the same as those in Table \ref{tab:params1}.
    }
    \label{fig:IncohRatCharts}
 \end{figure*}

\textit{Weak Coupling --- }
The difference between power extracted by an incoherent trap per site from the dimer, trimer, and quadmer ring systems in the weakly coupled regime is captured by Fig.~\ref{fig:IncohRatCharts}a.
In the top panel of this figure we show which antenna system generates the most power per site for varying extraction and trap decay rates,  $\Gamma_{X}$ and $\Gamma_{t}^{\downarrow}$, whereas the middle and bottom panels capture the difference between the power extracted per site by the trimer with the dimer and quadmer systems, $\Delta  \bar{P}_{32}$ and $\Delta \bar{P}_{34}$, respectively.

At fast extraction rates all excitations are moved to the trap without vibrational relaxation or radiative decay playing a major role and, as is known from Refs.~\cite{Fruchtman16, Higgins17}, larger antennae with interacting dipoles offer no advantage over individually absorbing chromophores. 
We attribute the vertical trough in the middle panel of Fig.~\ref{fig:IncohRatCharts}a to be the result of a balanced light-harvesting cycle, produced when the trap decay rate is equal to that of the optical rate -- this is the optimal model of conventional extraction.
However, for $\Gamma_{X} \geq \Gamma^{\downarrow}_{t}$ the trap will be saturated and the power that can be extracted reduced.
Therefore, because the current on the trap increases with $N$, the quadmer is outperformed by the trimer in this region, and the trimer outperformed by the dimer.

At slower extraction rates, we encounter a bottleneck in the light-harvesting cycle \cite{Higgins17}.
Unlike for the case of the dimer, this bottleneck can be partially mitigated by larger ring systems through optical ratcheting.
The relationship between the rates which constitute the aforementioned bottleneck is visibly present in the horizontal peak across the middle panel of Fig.~\ref{fig:IncohRatCharts}, where the dimer performs worse than the trimer ring system.
In this region, excitations are held on the dimer ring/antenna without being extracted by the trap whilst another photon arrives -- the opportunity to harvest more power is thereby missed. 
The trimer and quadmer, on the other hand, can continue to extract through ratcheting.
The ratcheting advantage of these larger ring systems is thereby at its optimum across this horizontal peak.
The advantage provided by trimer and quadmer systems over the dimer highlights the important difference between simple dark state protection and optical ratcheting.
The performance improvement of the quadmer over the trimer is partly a result of our approach; because the ring radius is fixed the F\"orster coupling between dipoles is stronger for the quadmer than the trimer.
We then find a larger separation of energy eigenstates, that follows from the increased dipole density, increasing the transition energy, thereby increasing the energy of absorbed photons, and increasing the power harvested.
In Appendix~\ref{app:toptimal}  we investigate an approach where, instead of the ring radius, the \textit{nearest-neighbour separation} is fixed.
In this case we find that the trimer ring antenna outperforms its quadmer counterpart.
Despite the increases in coupling strength, the results for the pentamer presented here demonstrate the diminishing returns with regards to power extracted per site:
As the number of antenna dipoles increases so does the current, and the trap thereby saturates.
It is noteworthy that a second tier of ratcheting is possible for $N\geq5$ \cite{Higgins17} but requires a more complex interplay of rates \cite{BrownThesis}, something unexplored here as it does not seem beneficial under relatively dilute solar irradiation.

\begin{figure*}[htbp]
    \includegraphics[height=11 cm, angle=0]{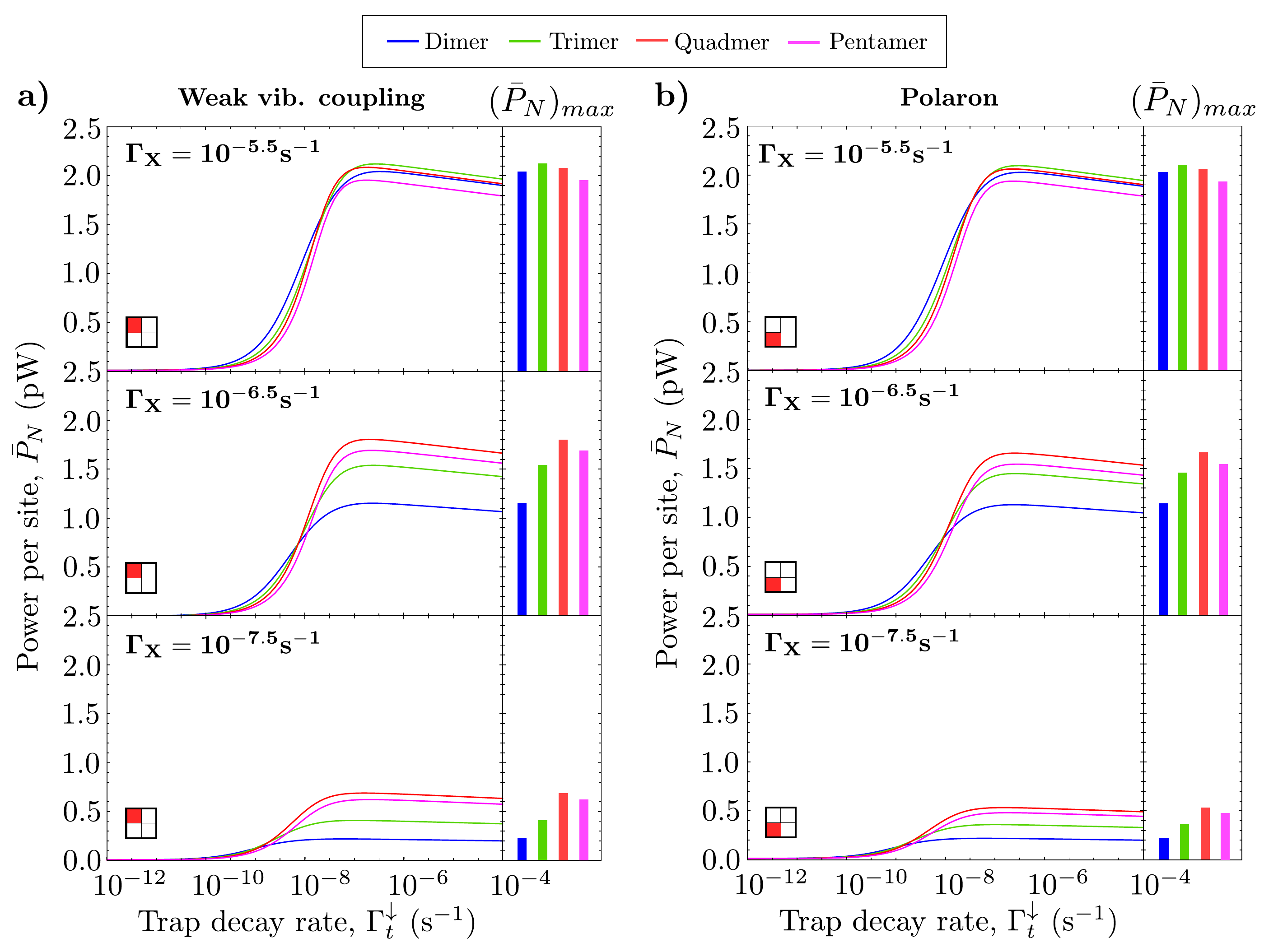}
    \caption{Power generated per site by an incoherent trap at varying decay rates $\Gamma_{t}^{\downarrow}$ for different values of $\Gamma_{X}$  in the weakly coupled and polaron regimes on the left and right hand side respectively.
    The values of $\Gamma_{X}$ are chosen across a range where the amount of power extracted by antennae is substantial whilst the hierarchy changes.
    Excluding those labelled, parameters are the same as those in Table \ref{tab:params1}.}
    \label{fig:incohpowslice}
 \end{figure*}

The absolute power per site extracted from these systems is shown by Fig.~\ref{fig:incohpowslice}a for three selected values of $\Gamma_X$ across the horizontal region where the ratcheting advantage is at its greatest.
Here, the increases in power harvested through the ratcheting mechanism are clearer: We observe that the trimer is the best antenna at `fast' extraction rates, $\Gamma_{X}=10^{-5.5}$~s\textsuperscript{$-1$}, demonstrating a slight ratcheting advantage.
This advantage becomes more evident at slower extraction rates, $\Gamma_{X}=10^{-6.5}$~s\textsuperscript{$-1$} and $\Gamma_{X}=10^{-7.5}$~s\textsuperscript{$-1$}, where we observe the quadmer, also utilising optical ratcheting, outperforming both the trimer and the dimer.
The pentamer trails behind the quadmer in each plot, demonstrating that ratcheting is less effective for antenna systems larger than the quadmer.

\textit{Polaron frame --- }
For extraction in the polaron frame, the spectral density described by Eqn.~\eqref{eq:specdens2} is used with $\lambda = 5$~meV and $\omega_{c}=90$~meV as in Ref.~\cite{Sowa17}.
Our choice corresponds to a relatively modest reorganisation energy of 5~meV with a resulting $B\approx0.98$ still close to unity (as is expected to be beneficial \cite{Rouse19}), whilst the vibrational spectrum is broad enough to support efficient phonon-assisted transitions across the entire width of the relevant excitation manifold(s). 
We repeat the analysis performed previously, the difference between power extracted by an incoherent trap per site from the dimer, trimer, and quadmer ring systems in the polaron frame is captured by the Fig.~\ref{fig:IncohRatCharts}b.

The same features are visible in Fig.~\ref{fig:IncohRatCharts}b, where we observe a vertical trough in the middle panel, identifying the region in which an optimal light-harvesting cycle is established in the dimer, whilst the horizontal peak of ratcheting persists.
Noticeably, the trimer broadly performs less well against the dimer in the polaron limit.
The introduction of more strongly coupled vibrational environments means that the ratcheting mechanism is undermined by the reduction of collective optical effects through the introduction of $P_{vib}$ in the optical rates.
Ratchet states which were previously totally dark can now experience optical absorption and (re-)emission processes, whilst emission rates associated with the bright states are similarly reduced \cite{Rouse19}.
This thereby limits the performance of ratcheting, which relies on absorbing a photon whilst another excitation is safely stored.
However, as long as the relaxation process is less likely to occur than excitation then ratcheting is still achievable and thus we see the ratcheting advantage persist.

The differences between the quadmer and the trimer are somewhat diminished in this regime, as can be seen in the bottom panel of Fig.~\ref{fig:IncohRatCharts}b.
Notably, the performance advantage of the trimer over the quadmer increases in the `fast' extraction regime. 
The effective rescaling of dipole-dipole coupling terms in the polaron limit captures a reduction in collective optical effects, the difference between power per site in general can therefore be expected to decrease.

Again, we show the absolute power extracted per site for three selected values of $\Gamma_X$ across the ratcheting region in Fig.~\ref{fig:incohpowslice}b.
We observe that the power extracted per site is largely the same for the first two `fast' extraction rates, $\Gamma_{X}=10^{-5.5}$~s\textsuperscript{$-1$} and $\Gamma_{X}=10^{-6.5}$~s\textsuperscript{$-1$}, with the power extracted by the quadmer and pentamer decreasing more substantially than the trimer.
This same trend is clearer at slower extraction rates,
$\Gamma_{X}=10^{-7.5}$~s\textsuperscript{$-1$}.
The most notable feature is the consistency of behaviour across this region, demonstrating the robustness of ratcheting.

\begin{figure}[htbp]
    \includegraphics[height=8.0 cm, angle=0]{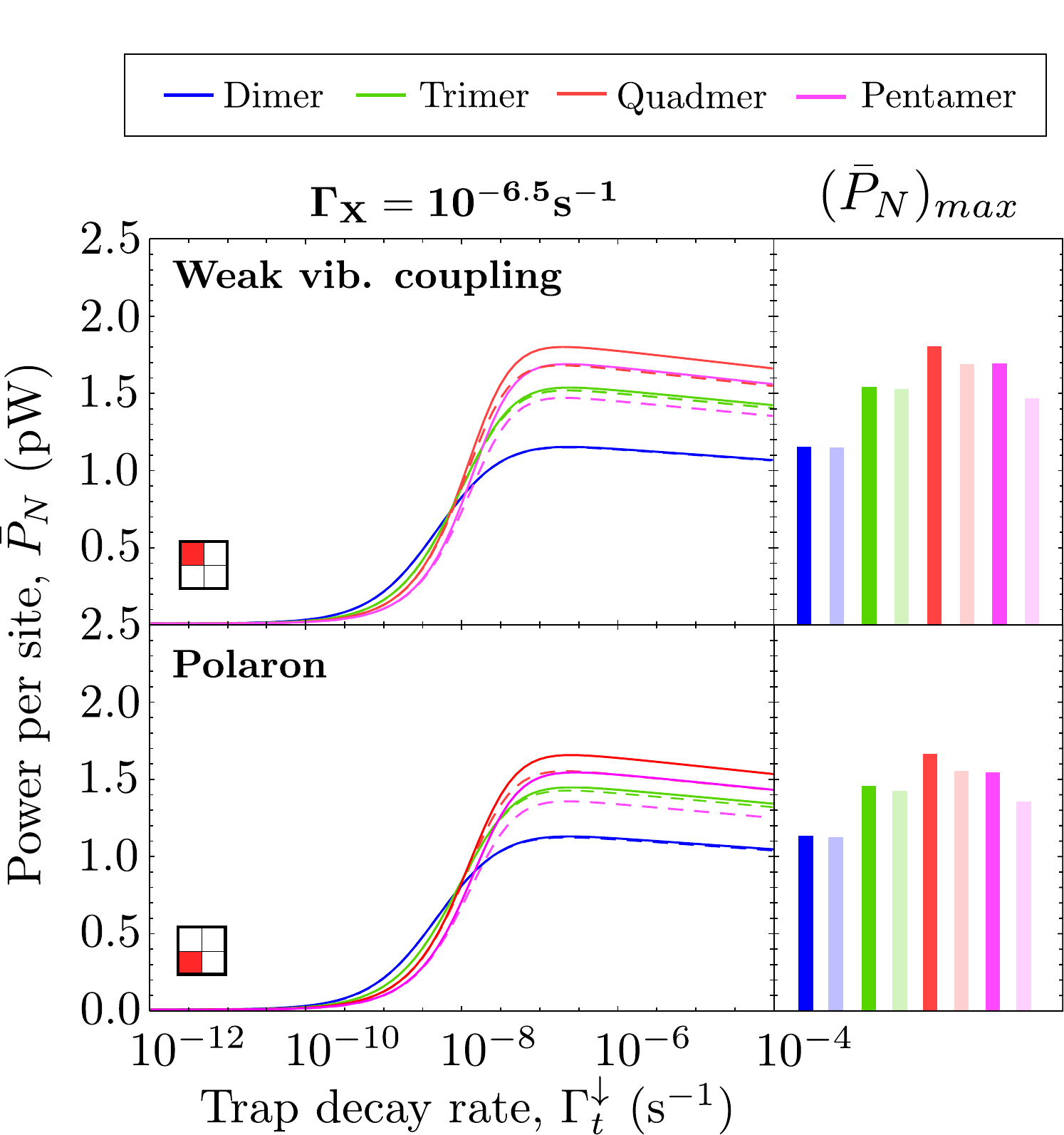}
    \abovecaptionskip=0pt
    \caption{Power generated per site by an incoherent trap at varying decay rates $\Gamma_{t}^{\downarrow}$ for $\Gamma_{X}=10^{-6.5}$~s\textsuperscript{$-1$} in the weakly coupled and polaron regimes on the top and bottom panel respectively with and without EEA at a rate of $\Gamma_{EEA}=10\kappa_{opt}$ shown by solid and dashed lines (saturated and faint bars), respectively).
    Excluding those labelled, parameters are the same as those in Table \ref{tab:params1}.}
    \label{fig:incoh_eea}
 \end{figure}

We now analyse the effect of EEA upon ratcheting. To this end, we consider the extraction rate $\Gamma_{X}=10^{-6.5}$~s\textsuperscript{$-1$} at which the ratcheting effect is strongly pronounced and implement a relatively aggressive EEA rate that exceeds the (single dipole) spontaneous decay rate by a factor of ten, $\Gamma_{EEA}=10\kappa_{opt}$. Figure~\ref{fig:incoh_eea} compares the pertinent results from Fig.~\ref{fig:incohpowslice} against those obtained when EEA is included. As expected, this shows that the dimer is robust against EEA, reflecting the fact that the dimer's dark state protection (or single tier ratcheting) does not rely on doubly-excited states. Surprisingly, the trimer is almost equally robust whilst \textit{still} being able to take advantage of a second tier of ratcheting. However, the performance of the quadmer and pentamer is significantly diminished.
We provide a more extensive investigation of relative power extracted from these ring systems across a broader range of EEA rates in Appendix~\ref{app:eea}.

\subsection{Coherent extraction}
\label{sec:rescoh}

A realistic description of extraction necessarily requires physical interaction. 
It is with this motivation that we extend the dipole-dipole couplings described by Eqn.~\eqref{eq:fcoup} to the trap, and thereby establish a physically well-founded extraction mechanism.
As such, we move to a regime where the trap is introduced as an additional physical dipole that is coherently coupled to the antenna.
Here exciton transport between antenna dipoles and the trap is mediated by F\"orster interactions.
We shall vary the trap energy, $\omega_{t}$, and distance from the ring to understand this model of extraction.

\begin{figure}[htbp]
    \includegraphics[height=8.0 cm, angle=0]{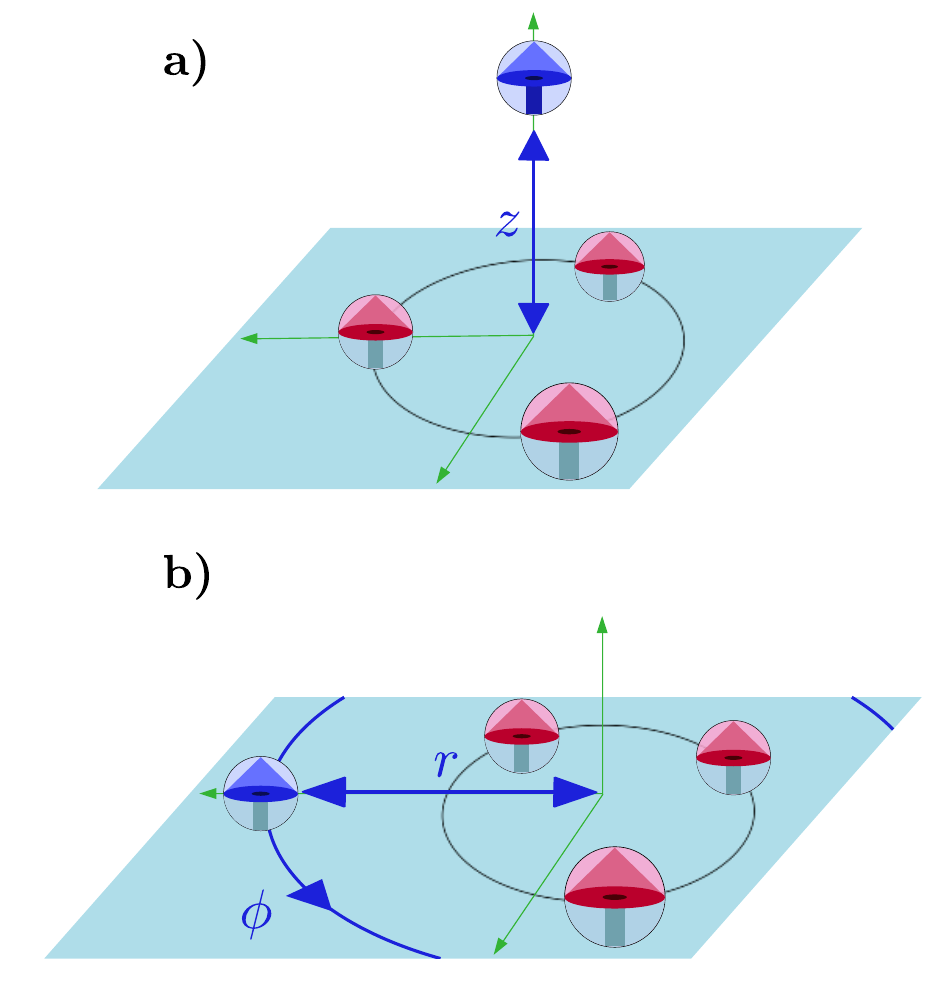}
    \abovecaptionskip=0pt
    \caption{Different geometrical arrangements of ring dipoles (red) and trap dipole (blue); the blue arrows indicate the direction of trap displacement. \textbf{a)} shows the trap positioned perpendicular to the plane of the ring and equidistant from all sites whereas in \textbf{b)} the trap is placed in the plane of the ring, parallel, and rotated around it for numerical optimisation over the polar angle $\phi$.}
    \label{fig:Alignment}
 \end{figure}

\begin{figure}[htbp]
    \includegraphics[height=4.0 cm, angle=0]{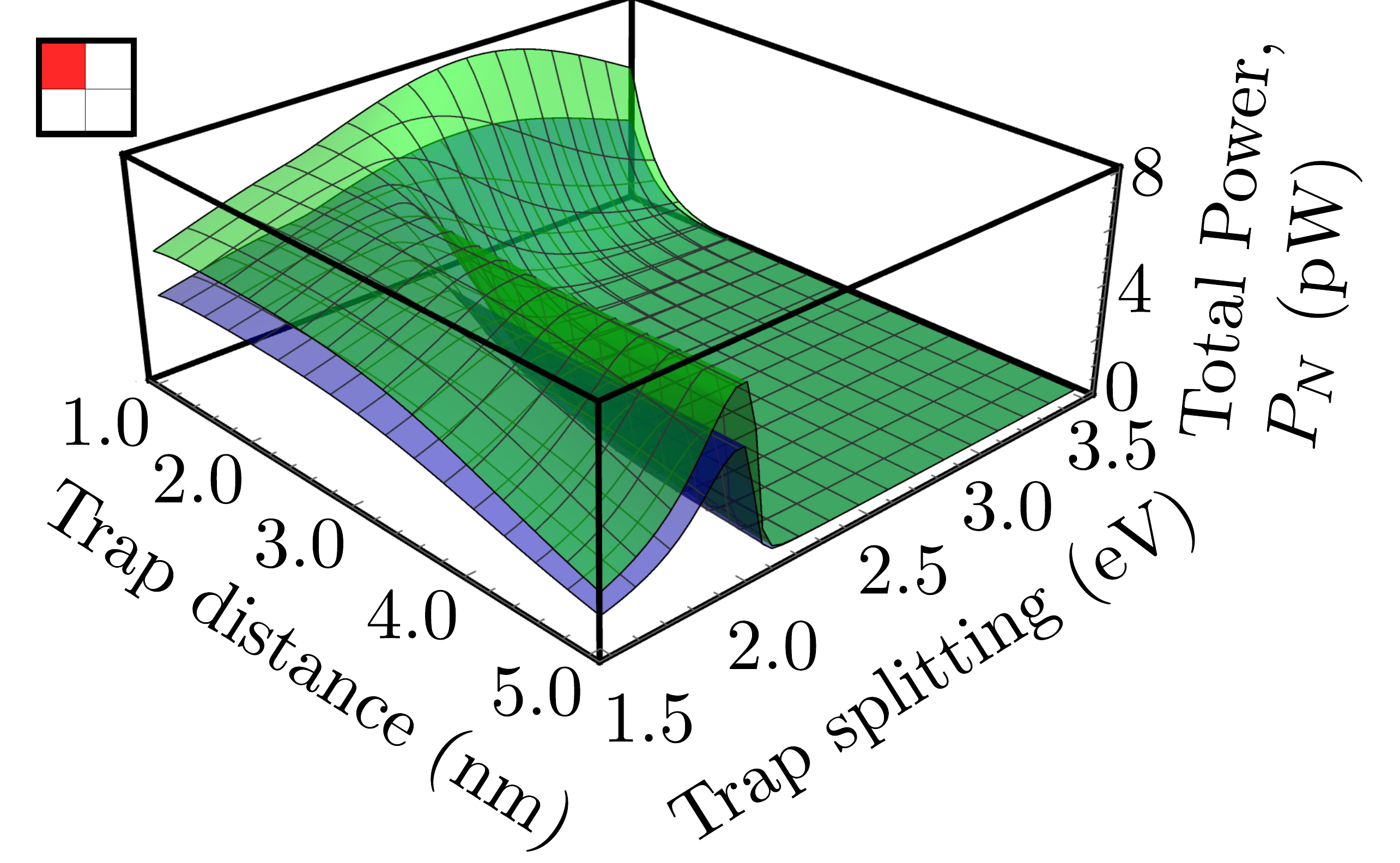}
    \abovecaptionskip=0pt
    \caption{Maximum power extracted \textit{per antenna} from a trimer (green) and dimer (blue) ring system by a coherent trap in the weakly coupled regime and orientation shown by Fig.~\ref{fig:Alignment}b.
    The distance of the trap from the antenna is varied alongside the trap splitting.
    We calculated the power over a broad range of trap decay rates $\Gamma^{\downarrow}_{t}=10^{-4}$ to $10^{-12}$ to determine the `maximum power' shown in the plot.
    Excluding those labelled and discussed, the parameters are the same as those in Table \ref{tab:params1}.
    }
    \label{fig:3dplot}
 \end{figure}

In the coherent model of extraction current and power per site are calculated to provide a measure of the ratcheting advantage in terms of the power extracted \textit{per site} and \textit{per antenna}.
These quantities are calculated at different distances from the ring in order to better understand how geometry affects coherent extraction.
We explore two different approaches towards the ring, as shown in Fig.~\ref{fig:Alignment}; the first is perpendicular to the plane of the ring and therefore equidistant from every ring dipole, the second has no such symmetry as the trap is placed parallel in the plane of the ring.
In the latter case, for a given $r$, we numerically optimise the value of $\phi$ for each ring system.
This is done to ensure fair comparison of the systems as the optimal geometric arrangement is different for the dimer and trimer.
Throughout, the trap's interaction with the optical bath is modelled such that all optical absorption and decay is from the ring system alone.

\begin{figure}[htbp]
    \includegraphics[width=0.45\textwidth, angle=0]{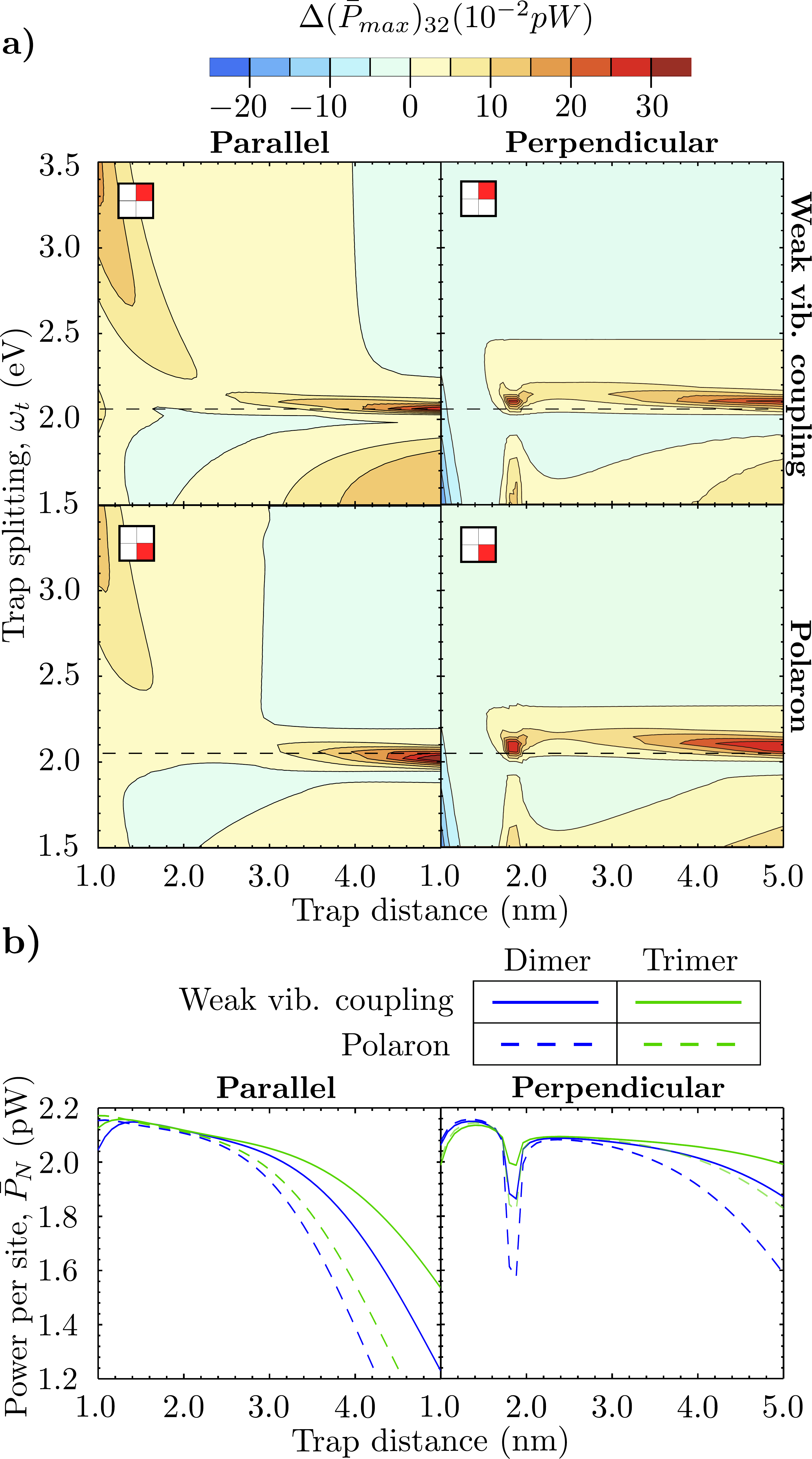}
    \abovecaptionskip=0pt
    \caption{
    A measure of the ratcheting advantage in extracting from a ring system with a coherent trap in the weakly coupled and polaron regimes for the two different orientations illustrated by Fig.~\ref{fig:Alignment}.
    In each instance presented the optimal trap decay rate and angle of orientation, $\phi$ (see Fig.~\ref{fig:Alignment}), has been chosen.
    The measure of the ratcheting advantage is illustrated by calculating the difference between the maximum power extracted \textit{per site} from a trimer and a dimer ring system.
    In the top figure, the left panels shows this measure calculated in the weak vibrational coupling and polaron regimes for the orientation shown by Fig.~\ref{fig:Alignment}b, whereas the right panels corresponds to Fig.~\ref{fig:Alignment}a.
    The explicit power harvested along the dashed line, capturing $\omega_{t}=2.05eV$, is shown in the bottom figure.
    Excluding those labelled and discussed, the parameters are the same as those in Table \ref{tab:params1}. 
    }
    \label{fig:CohRatCharts1}
 \end{figure}

In Fig.~\ref{fig:3dplot} we show the maximum power generated by a coherent trap across a range of trap splittings and distances for both the dimer and trimer ring systems \textit{per antenna} when the trap is in an orientation described by Fig.~\ref{fig:Alignment}b. This plot is for weak vibrational coupling and the maximum value is calculated by sweeping across different values for the trap decay rate $\Gamma^{\downarrow}_{t}$ between $10^{-12}$ and $10^{-4}$.
We observe that the trimer antenna invariably outperforms the dimer, albeit only very slightly so for trap splittings which exceed the energies of the antenna single excitation manifold, particularly when farther away.
The right hand side of Fig.~\ref{fig:CohRatCharts1}a shows the difference between a trimer and a dimer in the maximum power harvested \textit{per site} by a coherent trap in the same perpendicular configuration of Fig.~\ref{fig:Alignment}a now both in the weak coupling and polaron regimes.
The left hand side of Fig.~\ref{fig:CohRatCharts1}a below captures the same calculations for a parallel coherent trap, see Fig.~\ref{fig:Alignment}b.
Each panel focuses on the advantage of the trimer over a dimer antenna, $\Delta (\bar{P}_{max})_{32}$, i.e.~highlighting the advantage of an antenna that is capable of ratcheting in its simplest incarnation.
The explicit power extracted per site is captured by Fig.~\ref{fig:CohRatCharts1}b across a range of distances at $\omega_{t} = 2.05$~eV.

We now proceed to discuss the dependence of the ratcheting advantage on the geometrical trap configuration.
The perpendicular trap shows similar extraction behaviour between $1$ and $5$~nm; if the trap splitting is approximately degenerate with the lowest eigenenergy of the first excited manifold ($2$~eV) then ratcheting allows the trimer to outperform the dimer.
On the other hand, the parallel trap captures a very broad region wherein the trimer commonly outperforms the dimer at distances less than $4$~nm and up to larger separation for a suitably resonantly tuned trap.
The differences here demonstrate that the trimer advantage in near-field extraction is a result of the geometry of the trap in relation to the ring: In the parallel set-up, larger ring systems have more dipoles closer to the trap, whereas in the perpendicular set-up all dipoles remain equidistant.
The far-field case corresponds to the bottlenecked regime and we therefore observe ratcheting for which a tuning in the trap splitting is required.
A measure of the sensitivity of this tuning is investigated in Appendix~\ref{app:dext}.
We observe that in the case of a weakly coupled trap the far-field generates up to 20-60\% more power from a trimer ring than dimer in specific parameter configurations.
In line with our previous results, the advantage provided by optical ratcheting is sensitive to stronger vibrational coupling, but scope for improvement remains.

Lastly, we include a description of EEA for the trimer in the case of weak vibrational coupling. Once more, the trimer proves remarkably impervious to EEA, with a relative difference in extracted power of generally below $~1\%$. This further corroborates the picture of trimer antennae representing a highly robust ideal example of a ratcheting system.
Interestingly, EEA even appears to provide a minimal advantage when the trap splitting is large. 
This can be attributed to the EEA mechanism providing an efficient means of redistributing the small amount of population from the triply- and doubly-excited states down into the single excitation manifold, and thus to the transition that is most optimised for resonant extraction. However, this seeming advantage will be highly sensitive to the precise microscopic of the EEA process, and as such should be seen more as a quirk of our particular implementation than an easily realisable effect.

\begin{figure}[htbp]
    \includegraphics[height=9.0 cm, angle=0]{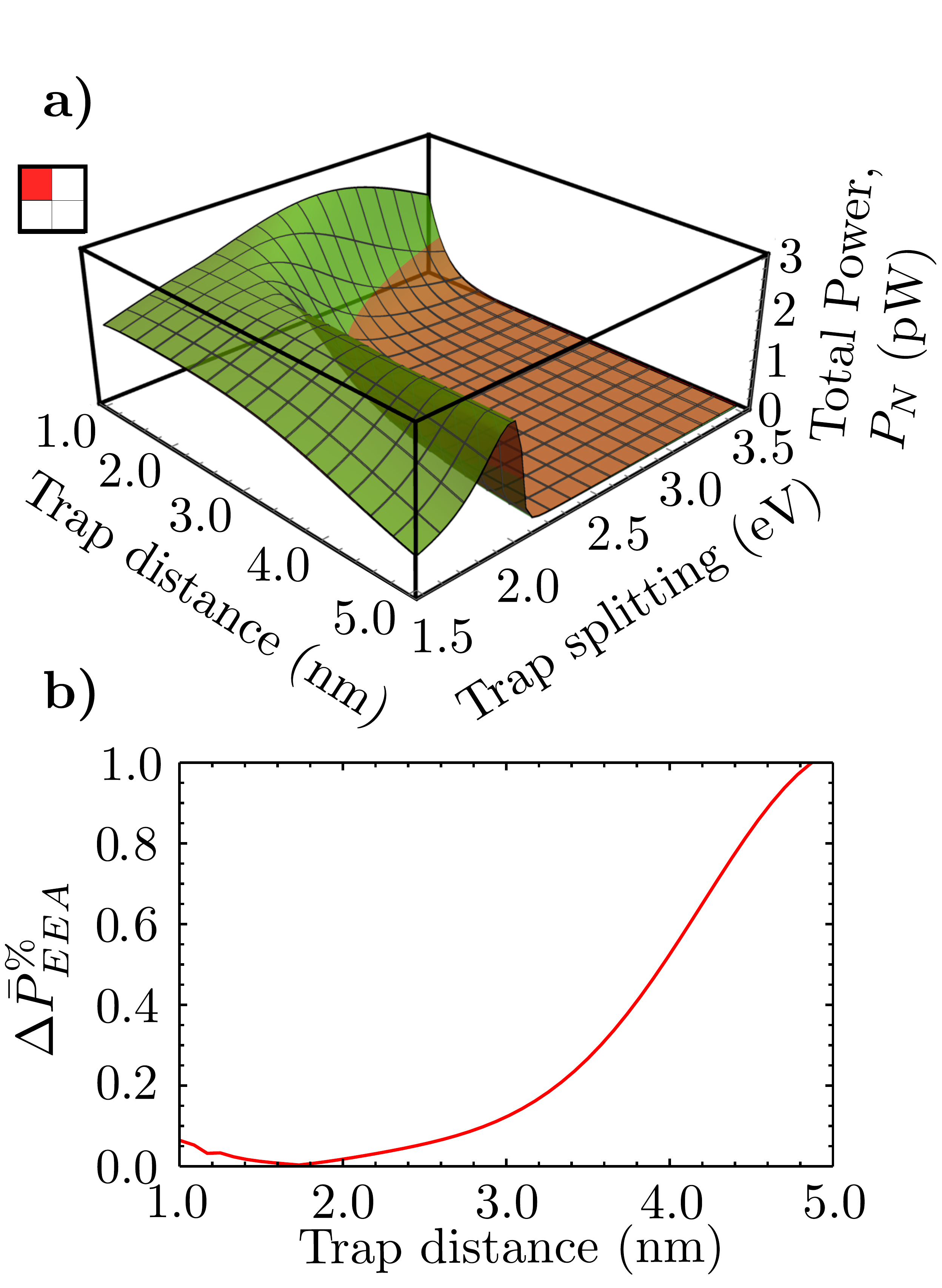}
    \abovecaptionskip=0pt
    \caption{In the top panel we show the power extracted \textit{per site} from a trimer ring system with and without EEA, in red and green respectively, by a coherent trap in the weakly coupled regime for the same set up as shown by Fig.~\ref{fig:3dplot}.
    In the bottom panel, the proportional difference between these two schemes at $\omega_{t} = 2.05$~eV is investigated.
    }
    \label{fig:3dplot_eea}
 \end{figure}

\section{Discussion} \label{sec:disc}

Optical ratcheting is a multi-excitation effect which relies on storing a single excitation in a state from which optical emission is unlikely, but further optical excitations are still possible.
The ratcheting process is beneficial when extraction from the ring to the trap is not too rapid, specifically when the energy conversion timescale is longer than an excitation would typically be expected to be held by a single dipole before re-emission. Importantly, we focus on the achievable power \textit{per absorbing dipole}. This means that any reported advantage indicates a situation where `the whole is greater than the sum of its parts', rather than just representing a trivial advantage of moving to a larger antenna system.

Through the incoherent trap model we demonstrated that the dimer system, which can only ratchet once to achieve dark state protection, is out-performed by trimer and quadmer systems, which are able to take advantage of two tiers of ratcheting when the extraction rate is less than both the trap decay rate and the rate of spontaneous emission, i.e.~ $\Gamma^{\downarrow}_{t} \gtrsim \Gamma_X$ and  $\Gamma_X \lesssim  \kappa_{opt}$.
The trimer ring emerges as the simplest and most robust antenna system, consistently extracting a near-optimal yield in the incoherent trap model.
We see this robustness in Appendix~\ref{app:toptimal}, where the ring radius is changed whilst the nearest-neighbour separation between dipoles changes, here the trimer outperforms all other ring arrangements.

The gains seen in the incoherent model are repeated in a more nuanced manner with the introduction of a coherent trap. 
Here, we considered the optimal extraction regime in two different coherent trap orientations in relation to the plane of the ring; perpendicular and parallel, as described by Fig.~\ref{fig:Alignment}.
Varying the trap energy and distance relative to the ring we find that for both the parallel and perpendicular orientations, the trap must be tuned with respect to trap energy if the trimer ratcheting advantage is to be observed.
Furthermore, the parallel trap demonstrates a broad trimer advantage for nearly all trap splittings and all trap distances below $4$~nm, however this is not the bottlenecked regime in which we expect to observe an improvement from ratcheting.
For both trap alignments ratcheting increasingly requires fine-tuning as the trap distance increases.
We see a reduced trimer advantage in the output generated by both the incoherently and coherently coupled traps following our shift to the polaron frame.

We have also analysed the effect of exciton-exciton annihilation (EEA) on the light-harvesting performance of ratcheting photocells.
In previous work, the effects of EEA upon ratcheting have been shown to be significant for weakly coupled quadmer systems with incoherent extraction when the annihilation rate exceeded the spontaneous emission rate by more than an order of magnitude \cite{Higgins17}.
Conversely, when EEA can be controlled and suppressed, to a low level, the benefit of ratcheting prevailed.
Through the analysis conducted in previous sections we reveal that
although quadmer ring systems are somewhat vulnerable to EEA, the corresponding trimer system proves remarkably robust.
This has been demonstrated in both the incoherent and coherent extraction models under a range of extraction rates.
This extends the potential for ratcheting to be a candidate for power and efficiency enhancement in realistic physical systems.

It is worth highlighting a final subtlety: the energetic spectrum of the ring antenna is known to depend on the number of constituent dipoles \cite{Hu18b}. Whilst our free-space optical spectral density only varies slowly with frequency, this will nonetheless slightly affect the precise values for the various emission and absorption rates, and thus introduce subtle differences when comparing ring antennae of different sizes. This is only a minor effect in the context of this study, but could become a very important factor for engineered photonic environments (such as an optical cavity or photonic band gap); however, such a scenario might also necessitate upgrading to a non-Markovian approach of modelling the light-matter interaction.

\section{Conclusion} \label{sec:conc}

In this Article, we have given a thorough analysis of the power extracted by an external trap coupled to several different ring systems, each composed of a varying number of dipoles.
In particular, we identify and demonstrate the performance improvement provided by a specific mechanism -- \textit{optical ratcheting}. 

The advantage of ratcheting is presented here for both incoherent and coherent traps in a number of different set-ups. 
Only for a coherent trap described in the polaron frame do we see the ratcheting mechanism constrained.
From these results we can therefore expect the experimental realisation of ratcheting in such systems to require fine-tuning and therefore be vulnerable to noise.

As we have not explicitly included disorder and non-radiative loss in our model, our findings describe a more idealised system than presented by most artifically engineered or naturally occurring molecular ring antennae.
Regardless, as argued in Ref.~\cite{Higgins17}, the ratcheting advantage \textit{as a proportion of power harvested} will remain.

Our work provides a broad foundation on which to build quantum light-harvesting devices that could realise this advantage and specifically that ratcheting offers significant improvements in power generation.
It would be interesting to elaborate upon the model outlined here, combining it with insights on driving efficient charge carrier separation \cite{Gelinas14, Rui20} to design a complete light-harvesting architecture, with the aim to facilitate probing and ultimately utilising experimental realisations of optical ratcheting.

\section*{Acknowledgements}

NW and EG acknowledge funding from the EPSRC grant no.~EP/T007214/1. WB acknowledges studentship funding from EPSRC under grant no.~EP/L015110/1.

\bibliographystyle{unsrt}
\bibliography{refs} 

\begin{thebibliography}{10}

\bibitem{Scholes11}
G.~D. Scholes, G.~R. Fleming, A.~Olaya-Castro, and R.~van Grondelle.
\newblock Lessons from nature about solar light harvesting.
\newblock {\em Nat. Chem.}, 3(10):763--774, 2011.

\bibitem{Alharbi15}
F.~H. Alharbi and S.~Kais.
\newblock Theoretical limits of photovoltaics efficiency and possible
  improvements by intuitive approaches learned from photosynthesis and quantum
  coherence.
\newblock {\em Renewable and Sustainable Energy Reviews}, 43:1073--1089, 2015.

\bibitem{Hu18}
Z.~Hu, G.~S. Engel, and S.~Kais.
\newblock Double-excitation manifold{'}s effect on exciton transfer dynamics
  and the efficiency of coherent light harvesting.
\newblock {\em Phys. Chem. Chem. Phys.}, 20:30032--30040, 2018.

\bibitem{Tomasi19}
S.~Tomasi, S.~Baghbanzadeh, S.~Rahimi-Keshari, and I.~Kassal.
\newblock Coherent and controllable enhancement of light-harvesting efficiency.
\newblock {\em Phys. Rev. A}, 100:043411, Oct 2019.

\bibitem{Blazquez19}
R.~S{\'a}ez-Bl{\'a}zquez, J.~Feist, E.~Romero, A.~I.
  Fern{\'a}ndez-Dom{\'\i}nguez, and F.~J. Garc{\'\i}a-Vidal.
\newblock Cavity-modified exciton dynamics in photosynthetic units.
\newblock {\em J. Phys. Chem. Lett.}, 10(15):4252--4258, 2019.
\newblock PMID: 31291109.

\bibitem{Shockley61}
W.~Shockley and H.~J. Queisser.
\newblock Detailed balance limit of efficiency of p‐n junction solar cells.
\newblock {\em Journal of Applied Physics}, 32(3):510--519, 1961.

\bibitem{Scully10}
M.~O. Scully.
\newblock Quantum photocell: Using quantum coherence to reduce radiative
  recombination and increase efficiency.
\newblock {\em Phys. Rev. Lett.}, 104:207701, May 2010.

\bibitem{Scully11}
M.~O. Scully, K.~R. Chapin, K.~E. Dorfman, and A.~Svidzinsky.
\newblock Quantum heat engine power can be increased by noise-induced
  coherence.
\newblock {\em Proceedings of the National Academy of Sciences},
  108(37):15097--15100, 2011.

\bibitem{Bittner14}
E.~R. Bittner and C.~Silva.
\newblock Noise-induced quantum coherence drives photo-carrier generation
  dynamics at polymeric semiconductor heterojunctions.
\newblock {\em Nature Communications}, 5(1):3119, 2014.

\bibitem{Trebbia22}
J.~B. Trebbia, Q.~Deplano, P.~Tamarat, and B.~Lounis.
\newblock Tailoring the superradiant and subradiant nature of two coherently
  coupled quantum emitters.
\newblock {\em Nature Communications}, 13(1):2962, 2022.

\bibitem{Andreani19}
L.~C. Andreani, A.~Bozzola, P.~Kowalczewski, M.~Liscidini, and L.~Redorici.
\newblock Silicon solar cells: toward the efficiency limits.
\newblock {\em Advances in Physics: X}, 4(1):1548305, 2019.

\bibitem{Tiedje84}
T.~Tiedje, E.~Yablonovitch, G.D. Cody, and B.G. Brooks.
\newblock Limiting efficiency of silicon solar cells.
\newblock {\em IEEE Transactions on Electron Devices}, 31(5):711--716, 1984.

\bibitem{Green84}
M.A. Green.
\newblock Limits on the open-circuit voltage and efficiency of silicon solar
  cells imposed by intrinsic auger processes.
\newblock {\em IEEE Transactions on Electron Devices}, 31(5):671--678, 1984.

\bibitem{Xue10}
J.~Xue.
\newblock Perspectives on organic photovoltaics.
\newblock {\em Polymer Reviews}, 50(4):411--419, 2010.

\bibitem{Zheng10}
Y.~Zheng and J.~Xue.
\newblock Organic photovoltaic cells based on molecular donor-acceptor
  heterojunctions.
\newblock {\em Polymer Reviews}, 50(4):420--453, 2010.

\bibitem{Cui19}
Y.~Cui, H.~Yao, L.~Hong, T.~Zhang, Y.~Tang, B.~Lin, Kaihu Xian, B.~Gao, C.~An,
  P.~Bi, W.~Ma, and J.~Hou.
\newblock {Organic photovoltaic cell with 17\% efficiency and superior
  processability}.
\newblock {\em National Science Review}, 7(7):1239--1246, 12 2019.

\bibitem{Rui20}
W.~Rui, Z.~Chunfeng, L.~Qian, Z.~Zhiguo, W.~Xiaoyong, and X.~Min.
\newblock Charge separation from an intra-moiety intermediate state in the
  high-performance pm6:y6 organic photovoltaic blend.
\newblock {\em Journal of the American Chemical Society}, 142(29):12751--12759,
  2020.
\newblock PMID: 32602706.

\bibitem{Cindy14}
C.~X. Zhao, A.~Y. Mao, and G.~Xu.
\newblock Junction capacitance and donor-acceptor interface of organic
  photovoltaics.
\newblock {\em Applied Physics Letters}, 105(6):063302, 2014.

\bibitem{Seok15}
Jeesoo Seok, Tae~Joo Shin, Sungmin Park, Changsoon Cho, Jung-Yong Lee,
  Du~Yeol~Ryu, Myung~Hwa Kim, and Kyungkon Kim.
\newblock Efficient organic photovoltaics utilizing nanoscale heterojunctions
  in sequentially deposited polymer/fullerene bilayer.
\newblock {\em Scientific Reports}, 5(1):8373, 2015.

\bibitem{Creatore13}
C.~Creatore, M.~A. Parker, S.~Emmott, and A.~W. Chin.
\newblock Efficient biologically inspired photocell enhanced by delocalized
  quantum states.
\newblock {\em Phys. Rev. Lett.}, 111:253601, Dec 2013.

\bibitem{Zhang15}
Y.~Zhang, S.~Oh, F.~H. Alharbi, G.~S. Engel, and S.~Kais.
\newblock Delocalized quantum states enhance photocell efficiency.
\newblock {\em Phys. Chem. Chem. Phys.}, 17:5743--5750, 2015.

\bibitem{Zhang16}
Y.~Zhang, A.~Wirthwein, F.~H. Alharbi, G.~S. Engel, and S.~Kais.
\newblock Dark states enhance the photocell power via phononic dissipation.
\newblock {\em Phys. Chem. Chem. Phys.}, 18:31845--31849, 2016.

\bibitem{Fruchtman16}
A.~Fruchtman, R.~G\'omez-Bombarelli, B.~W. Lovett, and E.~M. Gauger.
\newblock Photocell optimization using dark state protection.
\newblock {\em Phys. Rev. Lett.}, 117:203603, Nov 2016.

\bibitem{Rouse19}
D~M Rouse, E~M Gauger, and B~W Lovett.
\newblock Optimal power generation using dark states in dimers strongly coupled
  to their environment.
\newblock {\em New J. Phys.}, 21(6):063025, jun 2019.

\bibitem{Brown19}
W.~M. Brown and E.~M. Gauger.
\newblock Light harvesting with guide-slide superabsorbing condensed-matter
  nanostructures.
\newblock {\em J. Phys. Chem. Lett.}, 10(15):4323--4329, 2019.
\newblock PMID: 31251067.

\bibitem{Davidson20}
S.~Davidson, A.~Fruchtman, F.~A. Pollock, and E.~M. Gauger.
\newblock The dark side of energy transport along excitonic wires: On-site
  energy barriers facilitate efficient, vibrationally mediated transport
  through optically dark subspaces.
\newblock {\em J. Chem. Phys.}, 153(13):134701, 2020.

\bibitem{Mattioni21}
A.~Mattioni, F.~Caycedo-Soler, S.~F. Huelga, and M.~B. Plenio.
\newblock Design principles for long-range energy transfer at room temperature.
\newblock {\em Phys. Rev. X}, 11:041003, Oct 2021.

\bibitem{Davidson22}
S.~Davidson, F.~A. Pollock, and E.~M. Gauger.
\newblock Eliminating radiative losses in long-range exciton transport.
\newblock {\em PRX Quantum}, 2022.

\bibitem{MorenoCardoner19}
M.~Moreno-Cardoner, D.~Plankensteiner, L.~Ostermann, D.~E. Chang, and
  H.~Ritsch.
\newblock Subradiance-enhanced excitation transfer between dipole-coupled
  nanorings of quantum emitters.
\newblock {\em Phys. Rev. A}, 100:023806, Aug 2019.

\bibitem{Turschmann19}
P.~T\"urschmann, H.~Le~Jeannic, S.~F. Simonsen, H.~R. Haakh, S.~G\"otzinger,
  V.~Sandoghdar, P.~Lodahl, and N.~Rotenberg.
\newblock Coherent nonlinear optics of quantum emitters in nanophotonic
  waveguides.
\newblock {\em Nanophotonics}, 8(10):1641--1657, 2019.

\bibitem{Holzinger21}
R.~Holzinger, M.~Moreno-Cardoner, and H.~Ritsch.
\newblock Nanoscale continuous quantum light sources based on driven dipole
  emitter arrays.
\newblock {\em Applied Physics Letters}, 119(2):024002, 2021.

\bibitem{Masson2022}
S.~J. Masson and A.~Asenjo-Garcia.
\newblock Universality of dicke superradiance in arrays of quantum emitters.
\newblock {\em Nature Communications}, 13(1):2285, 2022.

\bibitem{Guerin16}
W.~Guerin, M.~O. Ara\'ujo, and R.~Kaiser.
\newblock Subradiance in a large cloud of cold atoms.
\newblock {\em Phys. Rev. Lett.}, 116:083601, Feb 2016.

\bibitem{Solano2017}
P.~Solano, P.~Barberis-Blostein, F.~K. Fatemi, L.~A. Orozco, and S.~L. Rolston.
\newblock Super-radiance reveals infinite-range dipole interactions through a
  nanofiber.
\newblock {\em Nature Communications}, 8(1):1857, 2017.

\bibitem{Holzinger20}
R.~Holzinger, D.~Plankensteiner, L.~Ostermann, and H.~Ritsch.
\newblock Nanoscale coherent light source.
\newblock {\em Phys. Rev. Lett.}, 124:253603, Jun 2020.

\bibitem{Holzinger22}
R.~Holzinger, S.~A. Oh, M.~Reitz, H.~Ritsch, and C.~Genes.
\newblock Cooperative subwavelength molecular quantum emitter arrays, 2022.

\bibitem{MorenoCardoner22}
M.~Moreno-Cardoner, R.~Holzinger, and H.~Ritsch.
\newblock Efficient nano-photonic antennas based on dark states in quantum
  emitter rings.
\newblock {\em Opt. Express}, 30(7):10779--10791, Mar 2022.

\bibitem{OSullivan11}
M.~C. O'Sullivan, J.~K. Sprafke, D.~V. Kondratuk, C.~Rinfray, T.~D.~W.
  Claridge, A.~Saywell, M.~O. Blunt, J.~N. O'Shea, P.~H. Beton, M.~Malfois, and
  H.~L. Anderson.
\newblock Vernier templating and synthesis of a 12-porphyrin nano-ring.
\newblock {\em Nature}, 469(7328):72--75, 2011.

\bibitem{Raymond08}
J.~E. Raymond, A.~Bhaskar, T.~Goodson, N.~Makiuchi, K.~Ogawa, and Y.~Kobuke.
\newblock Synthesis and two-photon absorption enhancement of porphyrin
  macrocycles.
\newblock {\em Journal of the American Chemical Society}, 130(51):17212--17213,
  2008.
\newblock PMID: 19053414.

\bibitem{Zheng16}
T.~Zheng, Z.~Cai, R.~Ho-Wu, S.~H. Yau, V.~Shaparov, T.~Goodson, and L.~Yu.
\newblock Synthesis of ladder-type thienoacenes and their electronic and
  optical properties.
\newblock {\em Journal of the American Chemical Society}, 138(3):868--875,
  2016.
\newblock PMID: 26720200.

\bibitem{Cai17}
Z.~Cai, R.~J. V{\'a}zquez, D.~Zhao, L.~Li, W.-Y. Lo, N.~Zhang, Q.~Wu,
  B.~Keller, A.~Eshun, N.~Abeyasinghe, H.~Banaszak-Holl, T.~Goodson, and L.~Yu.
\newblock Two photon absorption study of low-bandgap, fully conjugated perylene
  diimide-thienoacene-perylene diimide ladder-type molecules.
\newblock {\em Chemistry of Materials}, 29(16):6726--6732, 2017.

\bibitem{Richert17}
S.~Richert, J.~Cremers, I.~Kuprov, M.~D. Peeks, H.~L. Anderson, and C.~R.
  Timmel.
\newblock Constructive quantum interference in a bis-copper six-porphyrin
  nanoring.
\newblock {\em Nature Communications}, 8(1):14842, 2017.

\bibitem{Judd20}
C.~J. Judd, A.~S. Nizovtsev, R.~Plougmann, D.~V. Kondratuk, H.~L. Anderson,
  E.~Besley, and A.~Saywell.
\newblock Molecular quantum rings formed from a $\ensuremath{\pi}$-conjugated
  macrocycle.
\newblock {\em Phys. Rev. Lett.}, 125:206803, Nov 2020.

\bibitem{Chang83}
C.~K. Chang and I.~Abdalmuhdi.
\newblock Anthracene pillared cofacial diporphyrin.
\newblock {\em The Journal of Organic Chemistry}, 48(26):5388--5390, 1983.

\bibitem{Ouyang09}
Q.~Ouyang, Y.-Z. Zhu, C.-H. Zhang, K.-Q. Yan, Y.-C. Li, and J.-Y. Zheng.
\newblock An efficient pifa-mediated synthesis of fused diporphyrin and
  triply-singly interlacedly linked porphyrin array.
\newblock {\em Organic Letters}, 11(22):5266--5269, 2009.
\newblock PMID: 19835399.

\bibitem{Metselaar08}
G.~A. Metselaar, J.~K.~M. Sanders, and J.~de~Mendoza.
\newblock A self-assembled aluminium(iii) porphyrin cyclic trimer.
\newblock {\em Dalton Trans.}, pages 588--590, 2008.

\bibitem{GilRamirez10}
G.~Gil-Ram\'{i}rez, S.~D. Karlen, A.~Shundo, K.~Porfyrakis, Y.~Ito, G.~A.~D.
  Briggs, J.~J.~L. Morton, and H.~L. Anderson.
\newblock A cyclic porphyrin trimer as a receptor for fullerenes.
\newblock {\em Organic Letters}, 12(15):3544--3547, 2010.
\newblock PMID: 20597475.

\bibitem{Maeda20}
C.~Maeda, S.~Toyama, N.~Okada, K.~Takaishi, S.~Kang, D.~Kim, and T.~Ema.
\newblock Tetrameric and hexameric porphyrin nanorings: Template synthesis and
  photophysical properties.
\newblock {\em Journal of the American Chemical Society}, 142(37):15661--15666,
  2020.
\newblock PMID: 32847356.

\bibitem{Liu16}
P.~Liu, Y.~Hisamune, M.~D. Peeks, B.~Odell, J.~Q. Gong, L.~M. Herz, and H.~L.
  Anderson.
\newblock Synthesis of five-porphyrin nanorings by using ferrocene and
  corannulene templates.
\newblock {\em Angewandte Chemie International Edition}, 55(29):8358--8362,
  2016.

\bibitem{Yu18}
L.~Yu and J.~S. Lindsey.
\newblock Rational syntheses of cyclic hexameric porphyrin arrays for studies
  of self-assembling light-harvesting systems.
\newblock {\em J. Org. Chem.}, 66(22):7402--7419, 2001.
\newblock PMID: 11681955.

\bibitem{Hemmig16}
E.~A. Hemmig, C.~Creatore, B.~W\"{u}nsch, L.~Hecker, P.~Mair, M.~A. Parker,
  S.~Emmott, P.~Tinnefeld, U.~F. Keyser, and A.~W. Chin.
\newblock Programming light-harvesting efficiency using dna origami.
\newblock {\em Nano Letters}, 16(4):2369--2374, 2016.
\newblock PMID: 26906456.

\bibitem{Ketterer18}
P.~Ketterer, A.~N. Ananth, D.~S. Laman~Trip, A.~Mishra, E.~Bertosin, M.~Ganji,
  J.~van~der Torre, P.~Onck, H.~Dietz, and C.~Dekker.
\newblock Dna origami scaffold for studying intrinsically disordered proteins
  of the nuclear pore complex.
\newblock {\em Nature Communications}, 9(1):902, 2018.

\bibitem{Dey21}
S.~Dey, C.~Fan, K.~V. Gothelf, J.~Li, C.~Lin, L.~Liu, N.~Liu, M.~A.~D.
  Nijenhuis, B.~Sacc{\`a}, F.~C. Simmel, H.~Yan, and P.~Zhan.
\newblock Dna origami.
\newblock {\em Nature Reviews Methods Primers}, 1(1):13, 2021.

\bibitem{Higgins17}
K.~D.~B. Higgins, B.~W. Lovett, and E.~M. Gauger.
\newblock Quantum-enhanced capture of photons using optical ratchet states.
\newblock {\em J. Phys. Chem}, 121(38):20714--20719, 2017.

\bibitem{Mukai99}
K.~Mukai, S.~Abe, and H.~Sumi.
\newblock Theory of rapid excitation-energy transfer from b800 to
  optically-forbidden exciton states of b850 in the antenna system lh2 of
  photosynthetic purple bacteria.
\newblock {\em The Journal of Physical Chemistry B}, 103(29):6096--6102, 1999.

\bibitem{Scholes00}
G.~D. Scholes and G.~R. Fleming.
\newblock On the mechanism of light harvesting in photosynthetic purple
  bacteria: B800 to b850 energy transfer.
\newblock {\em The Journal of Physical Chemistry B}, 104(8):1854--1868, 2000.

\bibitem{Warshel86}
A.~Warshel and J.-K. Hwang.
\newblock Simulation of the dynamics of electron transfer reactions in polar
  solvents: Semiclassical trajectories and dispersed polaron approaches.
\newblock {\em The Journal of Chemical Physics}, 84(9):4938--4957, 1986.

\bibitem{Varada92}
G.~V. Varada and G.~S. Agarwal.
\newblock Two-photon resonance induced by the dipole-dipole interaction.
\newblock {\em Phys. Rev. A}, 45:6721--6729, May 1992.

\bibitem{Curutchet17}
C.~Curutchet and B.~Mennucci.
\newblock Quantum chemical studies of light harvesting.
\newblock {\em Chemical Reviews}, 117(2):294--343, 2017.
\newblock PMID: 26958698.

\bibitem{Savage96}
H.~Savage, M.~Cyrklaff, G.~Montoya, W.~K\"uhlbrandt, and I.~Sinning.
\newblock Two-dimensional structure of light harvesting complex ii (lhii) from
  the purple bacterium rhodovulum sulfidophilum and comparison with lhii from
  rhodopseudomonas acidophila.
\newblock {\em Structure}, 4(3):243--252, 1996.

\bibitem{Hu01}
X.~Hu, A.~Damjanovi{\'c}, T.~Ritz, and K.~Schulten.
\newblock Architecture and mechanism of the light-harvesting apparatus of
  purple bacteria.
\newblock {\em Proc. Natl. Acad. Sci. U.S.A}, 95(11):5935--5941, 1998.

\bibitem{Breuer02Book}
H.~P. Breuer and F.~Petruccione.
\newblock {\em The Theory of Open Quantum Systems}.
\newblock Oxford University Press, 2002.

\bibitem{WurfelBook}
P.~W\"urfel and U.~W\"urfel.
\newblock {\em Physics of Solar Cells: From Basic Principles to Advanced
  Concepts}.
\newblock John Wiley \& Sons, Hoboken, 2009.

\bibitem{Knox02}
R.~S. Knox, G.~J. Small, and S.~Mukamel.
\newblock Low-temperature zero phonon lineshapes with various brownian
  oscillator spectral densities.
\newblock {\em Chemical Physics}, 281(1):1--10, 2002.

\bibitem{Toutounji02}
M.~M. Toutounji and G.~J. Small.
\newblock The underdamped brownian oscillator model with ohmic dissipation:
  Applicability to low-temperature optical spectra.
\newblock {\em The Journal of Chemical Physics}, 117(8):3848--3855, 2002.

\bibitem{Fassioli12}
F.~Fassioli, A.~Olaya-Castro, and G.~D. Scholes.
\newblock Coherent energy transfer under incoherent light conditions.
\newblock {\em The Journal of Physical Chemistry Letters}, 3(21):3136--3142,
  2012.
\newblock PMID: 26296019.

\bibitem{Kreisbeck14}
C.~Kreisbeck, T.~Kramer, and A.~Aspuru-Guzik.
\newblock Scalable high-performance algorithm for the simulation of exciton
  dynamics. application to the light-harvesting complex ii in the presence of
  resonant vibrational modes.
\newblock {\em Journal of Chemical Theory and Computation}, 10(9):4045--4054,
  2014.
\newblock PMID: 26588548.

\bibitem{Gross82}
M.~Gross and S.~Haroche.
\newblock Superradiance: An essay on the theory of collective spontaneous
  emission.
\newblock {\em Physics Reports}, 93(5):301--396, 1982.

\bibitem{Nazir16}
A.~Nazir and D.~P.~S. McCutcheon.
\newblock Modelling exciton{\textendash}phonon interactions in optically driven
  quantum dots.
\newblock {\em J. Condens. Matter Phys.}, 28(10):103002, feb 2016.

\bibitem{Sowa17}
J.~K. Sowa, J.~A. Mol, G.~A.~D. Briggs, and E.~M. Gauger.
\newblock Environment-assisted quantum transport through single-molecule
  junctions.
\newblock {\em Phys. Chem. Chem. Phys.}, 19:29534--29539, 2017.

\bibitem{Rouse22}
D.~M. Rouse, E.~M. Gauger, and B.~W. Lovett.
\newblock Analytic expression for the optical exciton transition rates in the
  polaron frame.
\newblock {\em Phys. Rev. B}, 105:014302, Jan 2022.

\bibitem{Dorfman13}
K.~Dorfman, D.~Voronine, S.~Mukamel, and M.~O. Scully.
\newblock Photosynthetic reaction center as a quantum heat engine.
\newblock {\em Proceedings of the National Academy of Sciences of the United
  States of America}, 110, 01 2013.

\bibitem{Caruso12}
D.~Caruso and A~Troisi.
\newblock Long-range exciton dissociation in organic solar cells.
\newblock {\em Proc. Natl. Acad. Sci. U.S.A}, 109(34):13498--13502, 2012.

\bibitem{Blankenship14Book}
R.~E. Blankenship.
\newblock {\em Molecular mechanisms of photosynthesis}.
\newblock John Wiley \& Sons, 2014.

\bibitem{Gaididei80}
Y.~B. Gaididei and A.~I. Onipko.
\newblock A kinetic theory of incoherent exciton annihilation.
\newblock {\em Molecular Crystals and Liquid Crystals}, 62(3-4):213--235, 1980.

\bibitem{May14}
V.~May.
\newblock Kinetic theory of exciton–exciton annihilation.
\newblock {\em The Journal of Chemical Physics}, 140(5):054103, 2014.

\bibitem{Gelbwaser17}
D.~Gelbwaser-Klimovsky and A.~Aspuru-Guzik.
\newblock On thermodynamic inconsistencies in several photosynthetic and solar
  cell models and how to fix them.
\newblock {\em Chem. Sci.}, 8:1008--1014, 2017.

\bibitem{BrownThesis}
W.~M. Brown.
\newblock Vibrationally assisted collective optical effects, 2020.

\bibitem{Hu18b}
Z.~Hu, G.~S. Engel, and S.~Kais.
\newblock Connecting bright and dark states through accidental degeneracy
  caused by lack of symmetry.
\newblock {\em J. Chem. Phys.}, 148(20):204307, 2018.

\bibitem{Gelinas14}
S.~G\'{e}linas, A.~Rao, A.~Kumar, S.~L. Smith, A.~W. Chin, J.~Clark, T.~S.
  van~der Poll, G.~C. Bazan, and R.~H. Friend.
\newblock Ultrafast long-range charge separation in organic semiconductor
  photovoltaic diodes.
\newblock {\em Science}, 343(6170):512--516, 2014.

\end{thebibliography}

\appendix

\section{Results for electric currents} \label{sec:curr}

\begin{figure*}[htbp]
    \includegraphics[width=12.0 cm, angle=0]{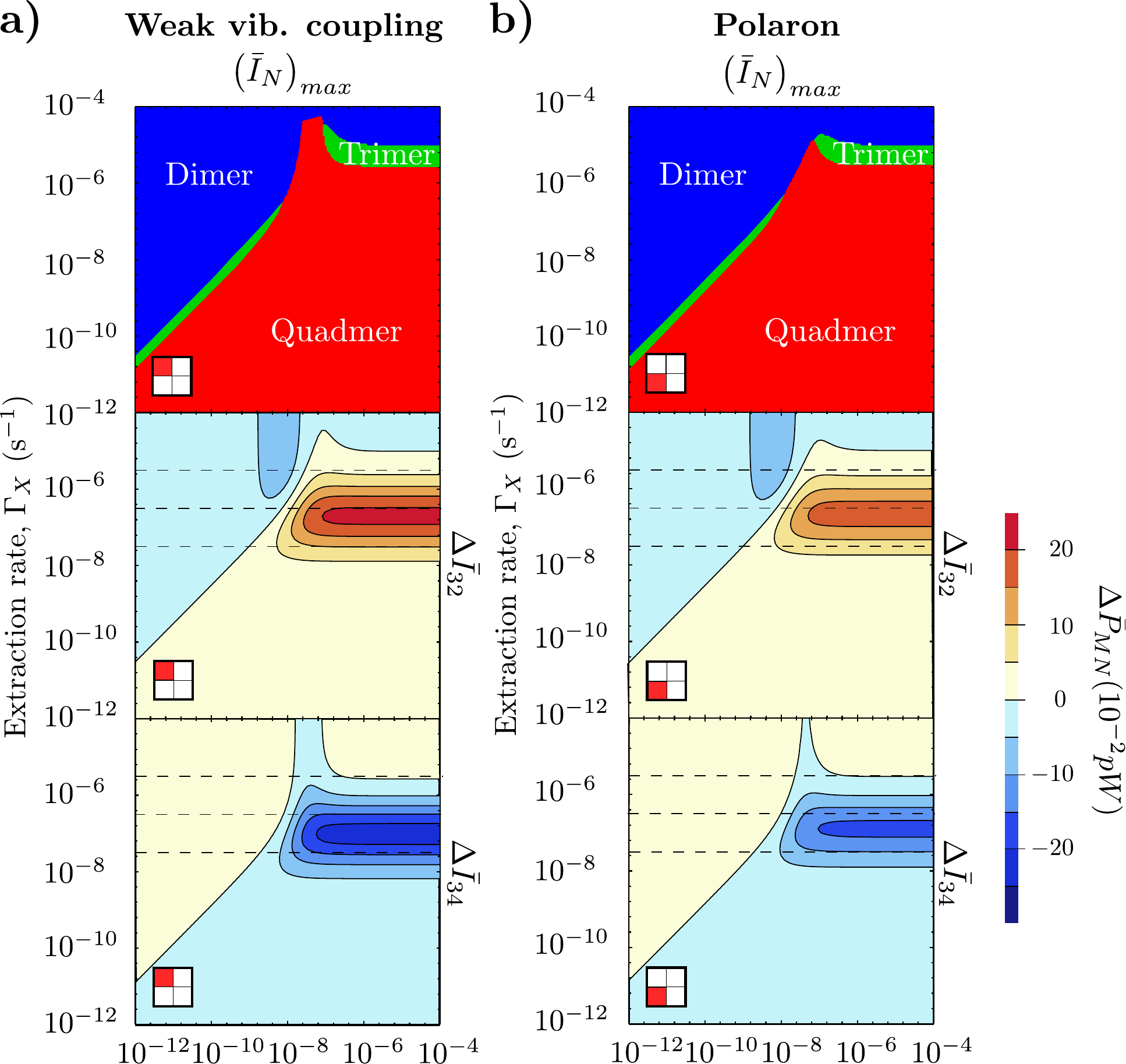}
    \abovecaptionskip=0pt
    \caption{Current generated from a ring system by an incoherent trap with a weak vibrational coupling on the left hand side and polaron frame on the right.
    In the top panel of both figures we highlight the optimal ring system, these differences are explicitly shown in the middle and right-hand side for the trimer relative to the dimer and quadmer, respectively.
    Values of $\Gamma_{X}$ used for later plots are indicated by dashed lines.
    Excluding those labelled, parameters are the same as those in Table \ref{tab:params1}.}
    \label{fig:curcont}
 \end{figure*}

\begin{figure*}[htbp]
    \includegraphics[height=11.0 cm, angle=0]{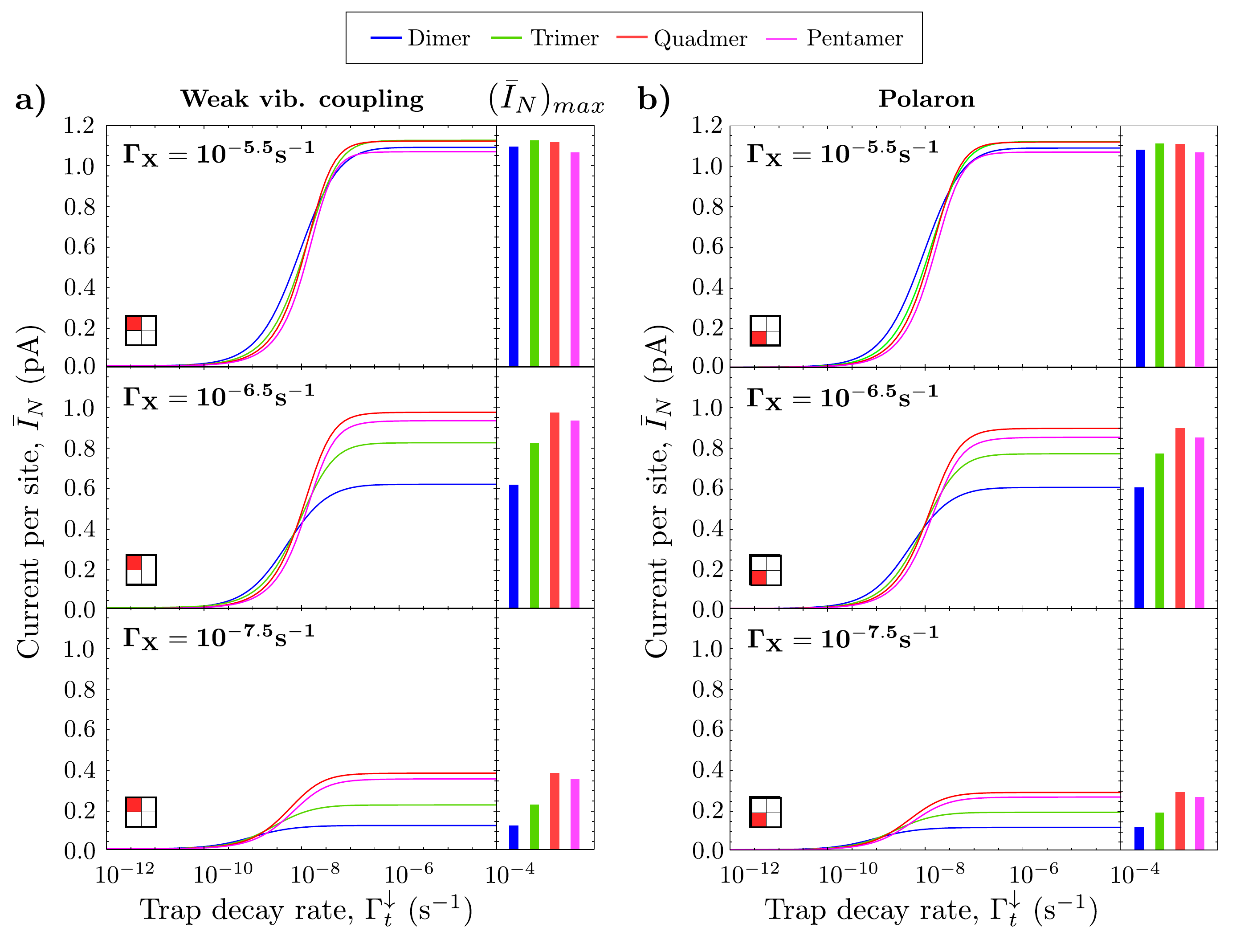}
    \abovecaptionskip=0pt
    \caption{Current generated per site by an incoherent trap at varying decay rates $\Gamma_{t}^{\downarrow}$ for different values of $\Gamma_{X}$.
    The values of $\Gamma_{X}$ are chosen across a range where the amount of power extracted by antennae is substantial whilst the hierarchy changes.
    Excluding those labelled, parameters are the same as those in Table \ref{tab:params1}.
    }
    \label{fig:curslice}
 \end{figure*}

\begin{figure*}[htbp]
    \includegraphics[width=0.5\textwidth, angle=0]{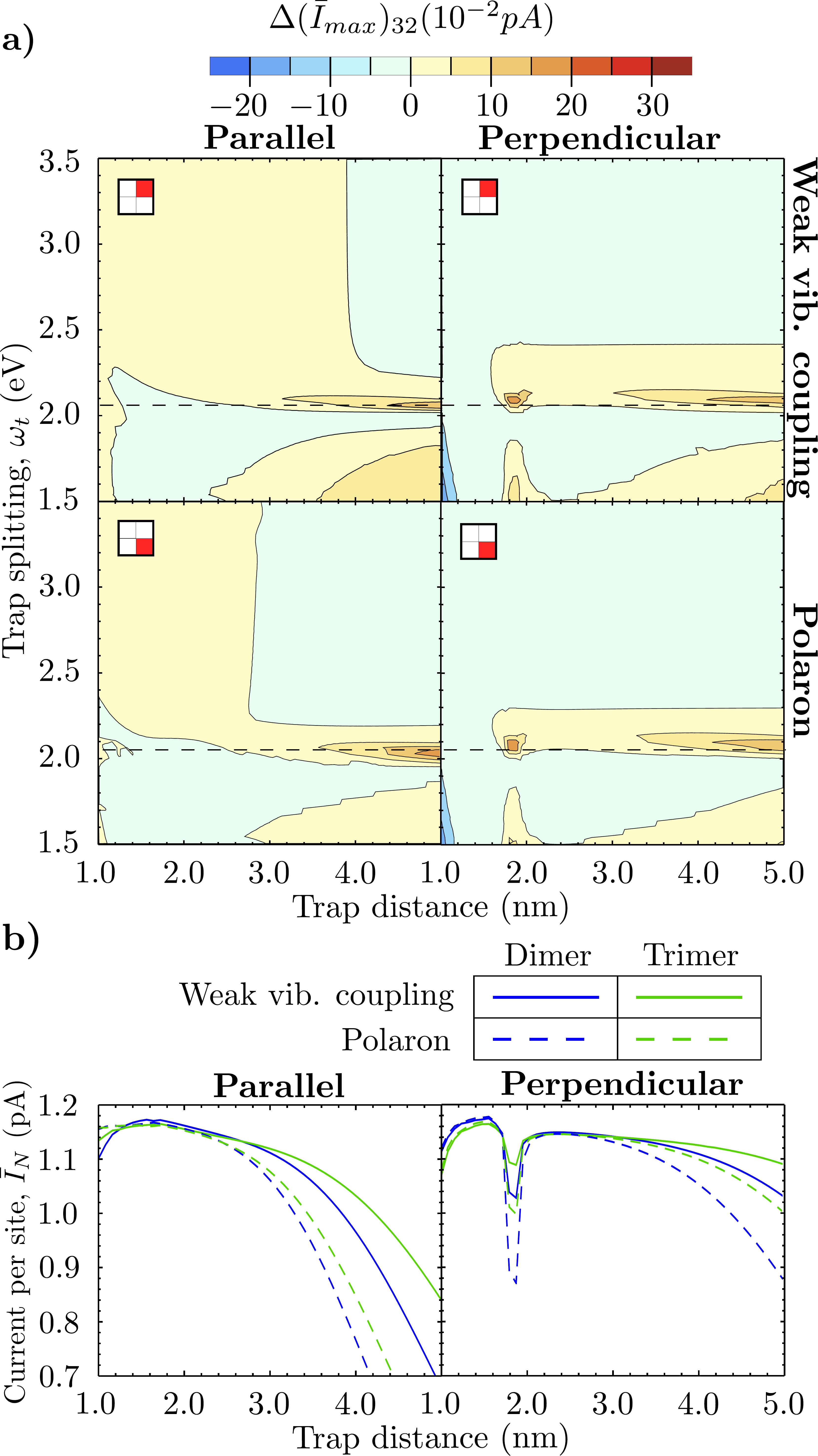}
    \abovecaptionskip=0pt
    \caption{
    A measure of the difference in current extracted from a ring system with a coherent trap in the weakly coupled and polaron regimes for the 2 different orientations illustrated by Fig.~\ref{fig:Alignment}.
    In each instance presented the optimal trap decay rate and angle of orientation, $\phi$ (see Fig.~\ref{fig:Alignment}), has been chosen.
    The difference calculated is between the current corresponding to the maximum power extracted \textit{per site} from a trimer and a dimer ring system.
    In the top figure, the left panels shows this measure calculated in the weak vibrational coupling and polaron regimes for the orientation shown by Fig.~\ref{fig:Alignment}b, whereas the right panels corresponds to Fig.~\ref{fig:Alignment}a.
    The explicit power harvested along the dashed line, capturing $\omega_{t}=2.05eV$, is shown in the bottom figure.
    Excluding those labelled and discussed, the parameters are the same as those in Table \ref{tab:params1}. 
    }
    \label{fig:currcohplots}
 \end{figure*}

Alongside the results for the power extracted by the incoherent trap in Section \ref{sec:resincoh} in Figs.~\ref{fig:curcont} and \ref{fig:curslice} we show the corresponding current values.
We see that these are broadly similar to the preceding figures.
The corresponding values for the current generated by the coherent trap, presented in Section \ref{sec:rescoh}, are shown in Fig.~\ref{fig:currcohplots}.
Again, these figures are broadly similar to those produced by the corresponding power calculations.

\section{Optimised trimer regime} \label{app:toptimal}
\begin{figure*}[htbp]
    \includegraphics[height=12.0cm, angle=0]{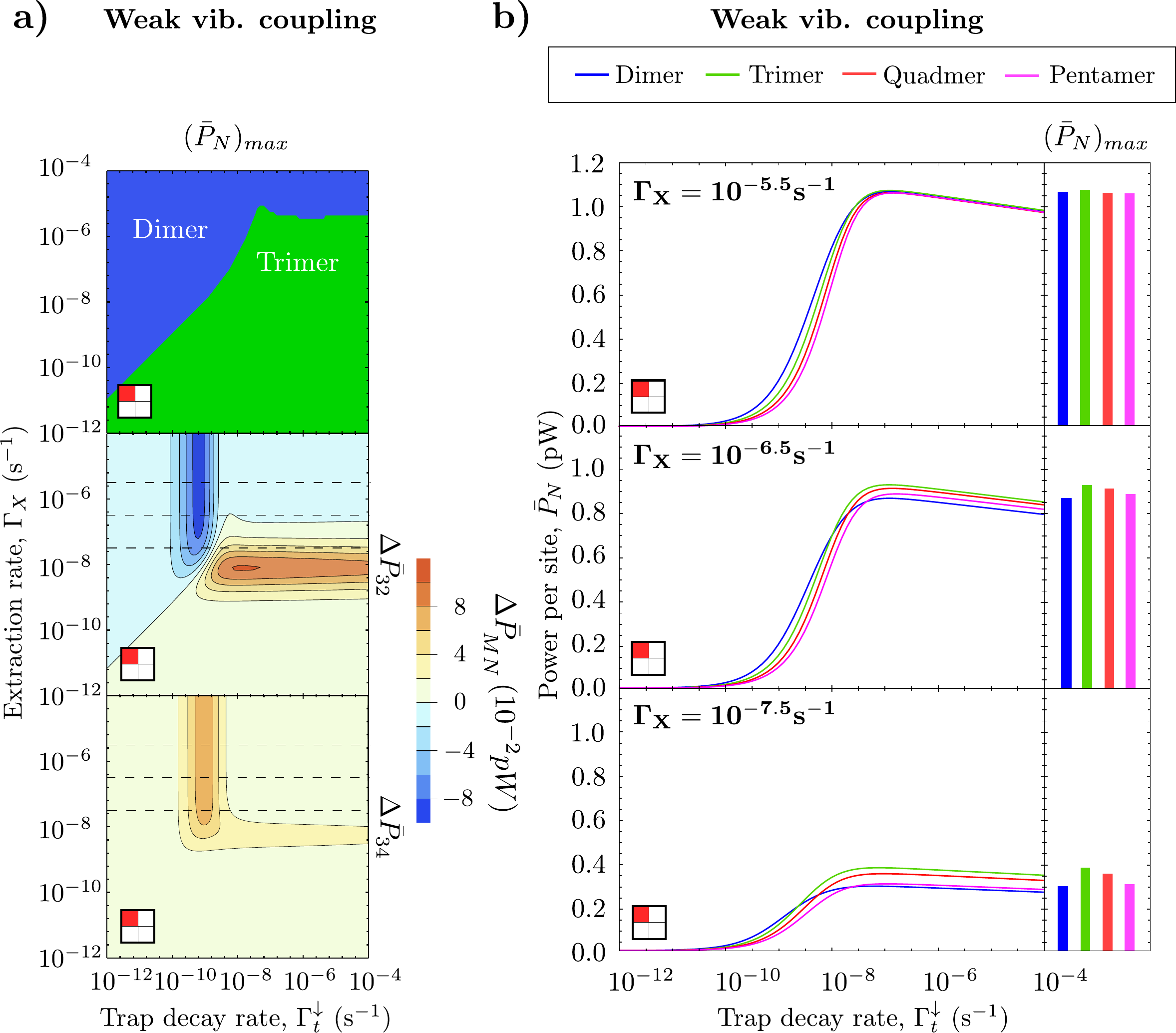}
    \abovecaptionskip=0pt
    \caption{
    Power generated from a ring system by an incoherent trap in the weakly coupled regimes such that the trimer dominates.
    We achieve such a scenario by setting a fixed nearest-neighbour distance, $s_{nn}$, of $2$~nm and a spontaneous emission rate $\kappa_{opt}$ of $5$~\si{ns}\textsuperscript{$-1$}.
    In the top panel of the left hand side we highlight the optimal ring system, while the difference in power generated by these systems is shown in the middle and bottom panels.    
    On the right hand side we show the explicit power generated from a ring system with an incoherent trap and a weak vibrational coupling at varying decay rates $\Gamma_{t}^{\downarrow}$ for different values of $\Gamma_{X}$.
    }
    \label{fig:incohplotsbesp}
 \end{figure*}

In Sections \ref{sec:resincoh} and \ref{sec:rescoh} we demonstrate the substantial ratcheting advantage presented by both trimer and quadmer ring systems over the dimer across a range of extraction and decay rates.
When comparing ring systems of fixed radius, those more densely populated by dipoles have larger separations between energy levels due to the increased coherent F\"{o}rster coupling.
As a result, for a ring system of radius $1.5$~nm we observe the quadmer outcompete the trimer in terms of power extraction through ratcheting.
Given the trimer is the simplest structure required for optical ratcheting in certain parameter regimes this system will provide a consistently optimal choice for power extraction as shown by Fig.~\ref{fig:incohplotsbesp}a. 
Fig.~\ref{fig:incohplotsbesp}b captures the absolute values for power extracted per site.
Notably, the ratcheting advantage during fast extraction is not as large as in the results previously discussed, rather it appears at slower extraction.

\section{Delocalised extraction} \label{app:dext}

\begin{figure*}[htbp]
    \includegraphics[height=5.5 cm, angle=0]{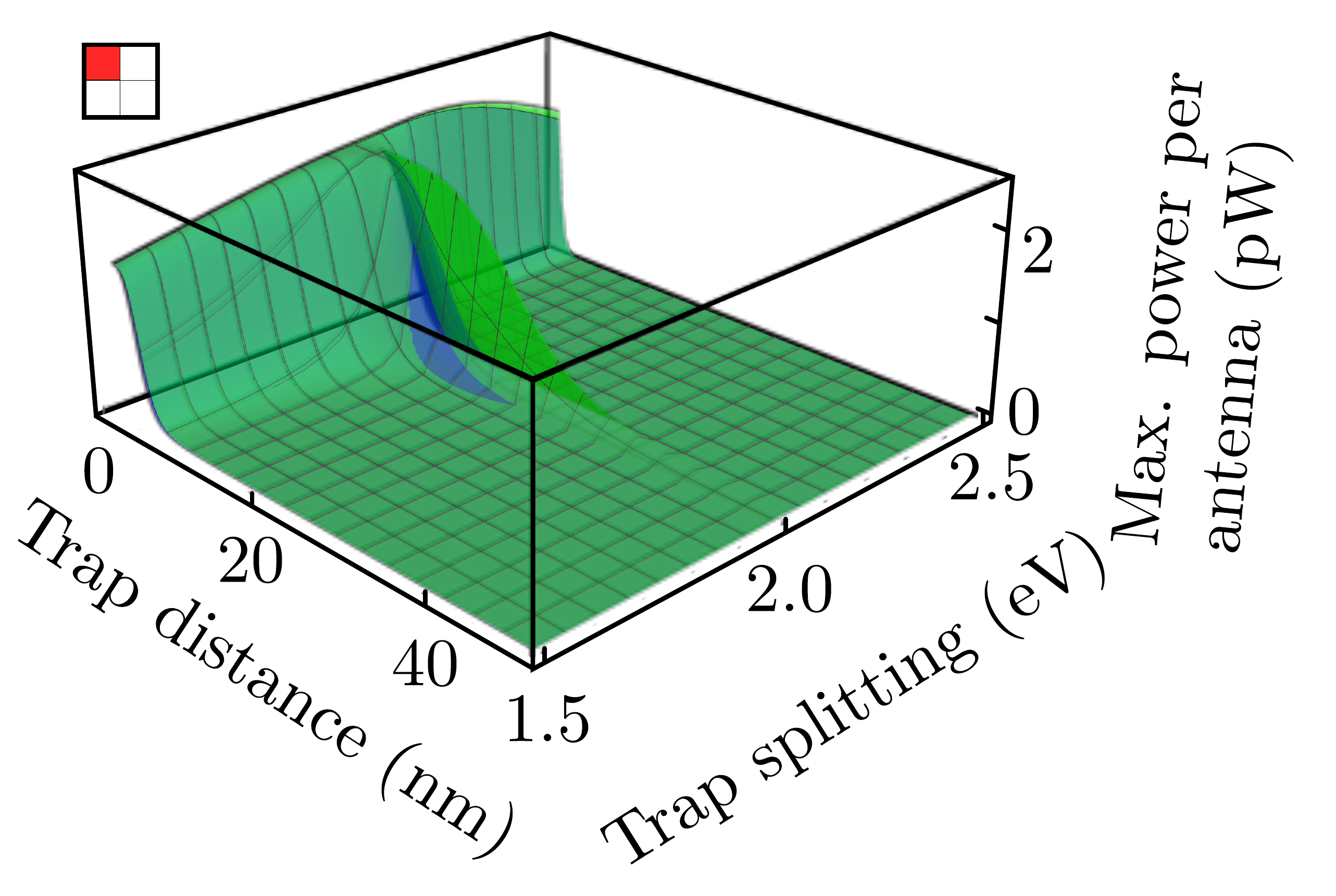}
    \abovecaptionskip=0pt
    \caption{Maximum power extracted \textit{per antenna} from a trimer (green) and dimer (blue) ring system by a weakly coupled coherent trap.
    Distance of the trap from the antenna is varied alongside the trap splitting.
    Excluding those labelled and discussed, the parameters are the same as those in Table \ref{tab:params1}.}
    \label{fig:delocalised3da} 
 \end{figure*}

\begin{figure*}[htbp]
    \includegraphics[height=4.5 cm, angle=0]{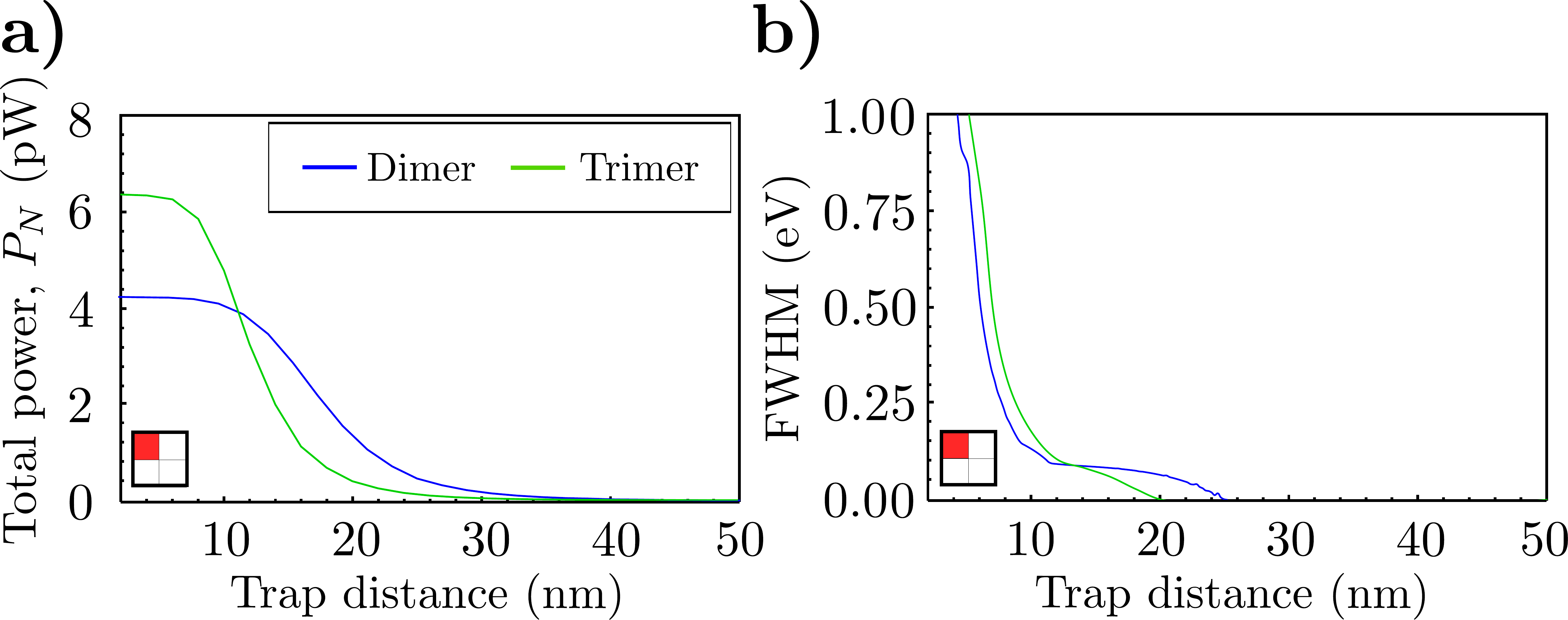}
    \abovecaptionskip=0pt
    \caption{On the left, maximum power extracted \textit{per antenna} from a trimer (green) and dimer (blue) ring system by a weakly coupled coherent trap at $\omega_{t}=2$~eV.
    The full width half maximum (FWHM) of this result is on the right.
    Distance of the trap from the antenna is varied alongside the trap splitting.
    For the values presented, the maximum power extracted is calculated from a range of trap decay rates; $\Gamma^{\downarrow}_{t}=10^{-4}$ to $10^{-12}$.
    Excluding those labelled and discussed, the parameters are the same as those in Table \ref{tab:params1}.}
    \label{fig:delocalised3db} 
 \end{figure*}

Both the dimer and trimer experience optimal extraction by the trap at particular trap splittings which become increasingly evident at larger distances.
The extraction shown by Fig.~\ref{fig:delocalised3da} and Fig.~\ref{fig:delocalised3db}a appears to show dramatic and sustained power output when the trap is tuned to the lowest eigenenergy of the ring system Hamiltonian.
When the trap and the ring dipoles are near-degenerate extraction is optimal.
However, the fine-tuning required for such extraction is incredibly sensitive to changes in trap splitting, and would be challenging to achieve practically, requiring an unrealistically stable trap -- as is demonstrated by the dramatic decrease of the FWHM shown in Fig.~\ref{fig:delocalised3db}b.

\section{Electron-Electron Annihilation} \label{app:eea}
\begin{figure*}[htbp]
    \includegraphics[width=1.0\textwidth, angle=0]{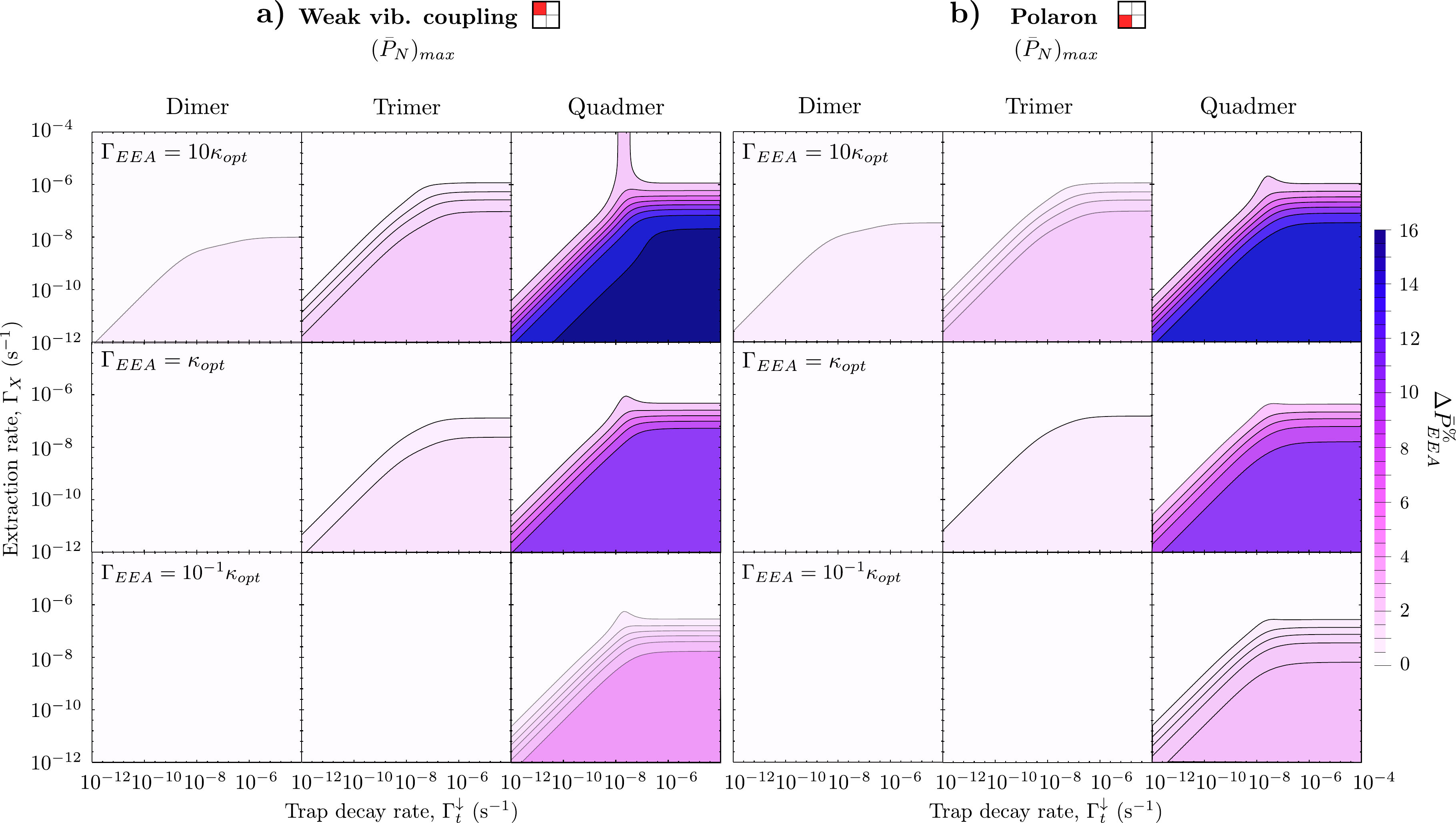}
    \abovecaptionskip=0pt
    \caption{
    In the presence of EEA, the proportional reduction in power generated from a ring system by an incoherent trap in the weakly coupled and polaron regimes on the left and right hand side, respectively.
    The proportional reduction in power is shown for the dimer, trimer, and quadmer systems.
    Excluding those labelled, parameters are the same as those in Table \ref{tab:params1}.
    }
    \label{fig:eea_NN}
 \end{figure*}

Beyond the examples provided in the manuscript, we explore the difference in power extracted by ring systems with and without EEA across a range of rates as shown by Fig.~\ref{fig:eea_NN}.

These results confirm the robustness of the trimer and dimer systems, which are relatively unaffected by EEA, with the proportional loss of power remaining under ~1\%.
In stark comparison, the power extracted from the quadmer is diminished significantly, with up to a 16\% decrease. This behaviour carries over into the polaron frame.

\end{document}